\documentstyle[times,graphics,astrobib,amssymb]{mn2e}

\newcommand{\be}{\begin{equation}} \newcommand{\e}{\end{equation}}
\newcommand{\bear}{\begin{eqnarray}} \newcommand{\ear}{\end{eqnarray}}
\newcommand{\nline}{\nonumber \\} \newcommand{\f}{\frac}
\newcommand{\de}{{\rm d}} 
\newcommand{\lya}{Ly$\alpha$} 

\begin{document}

\title[The topology of reionization] {Inside-out or  Outside-in: The
topology of reionization  in the photon-starved regime suggested by
\lya\ forest data} \author[Choudhury, Haehnelt \& Regan] {Tirthankar  Roy
Choudhury\thanks{E-mail: chou@ast.cam.ac.uk},~
Martin G. Haehnelt\thanks{E-mail: haehnelt@ast.cam.ac.uk} and
John Regan\thanks{E-mail: regan@ast.cam.ac.uk} \\ Institute of
Astronomy, Madingley Road, Cambridge CB3 0HA, UK}

\maketitle

\date{\today}

\begin{abstract} We use a set of semi-numerical simulations based on
Zel'dovich approximation, friends-of-friends algorithm and excursion
set formalism to generate reionization maps of high dynamic range with
a range of assumptions regarding the distribution and luminosity of
ionizing sources and the spatial distribution of sinks for the
ionizing radiation. We find that ignoring the inhomogeneous spatial
distribution  of regions of high gas density where recombinations are
important -- as is often done in studies of this kind -- can lead
to misleading conclusions regarding the topology of reionization,
especially if reionization occurs in the photon-starved regime
suggested by \lya\ forest data. The inhomogeneous spatial distribution
of recombinations significantly reduces the mean free
path of ionizing  photons and the typical size of coherently ionized
regions. Reionization proceeds then  much more as an outside-in
process. Low-density regions  far from ionizing sources become ionized
before regions of  high gas density not hosting sources of ionizing
radiation.  The spatial  distribution of
sinks of ionization radiation also significantly affects shape and
amplitude the power spectrum of fluctuations of 21cm emission. The
slope of the 21cm power spectrum as measured by upcoming 21cm
experiments should be able to  distinguish to what extent the topology
of reionization proceeds outside-in or inside-out while the evolution of
the amplitude of the   power spectrum with increasing ionized mass
fraction should be sensitive to the spatial distribution  and the
luminosity of ionizing sources. 
\end{abstract}
\section{Introduction}

The reionization of neutral hydrogen is an important milestone in the
evolution of the Universe. The epoch of reionization has received  a
major boost  of attention  recently due  to a series of observational
advances which suggest that the process is complex and that the
reionization of hydrogen extends  over wide redshift range from $6
\lesssim z \lesssim 15$ (for reviews see \citeNP{cf06a,fob06}).  We
are about  to enter  an  exciting phase as  planned 21cm observations are
expected to settle the  questions when and how the Universe was
reionized. It  thus timely  to develop more accurate and detailed
analytical and numerical models  in order to extract the maximum
information about the physical processes relevant  for reionization
from the expected large and complex  future data sets. 

Currently operating and upcoming low-frequency radio observations
(e.g., GMRT\footnote{http://www.gmrt.ncra.tifr.res.in/},
21CMA\footnote{http://web.phys.cmu.edu/~past/},
MWA\footnote{http://www.haystack.mit.edu/ast/arrays/mwa/},
LOFAR\footnote{http://www.lofar.org/},
SKA\footnote{http://www.skatelescope.org/})  of redshifted 21cm
emission of neutral hydrogen should also probe the  topology of the neutral
(or ionized) regions at high redshifts.   Unfortunately, modelling the
expected data sets is not straightforward  because of the dauntingly
wide range of physical scales involved and our  lack of knowledge of
many details of the relevant physical processes.

Full numerical simulations including radiative transfer effects  are
still computationally extremely challenging.  Modelling the  smallest
mass haloes contributing  to the  ionizing emissivity at early epochs
(with a total mass  $M \sim 10^8 M_{\odot}$) requires linear scales
$\lesssim 0.1$Mpc  while at the same time,  the size of the simulated
regions need to extend over  $100$ Mpc or more in order to probe the largest
coherently ionized regions in the final stages  of reionization.
Despite such challenging requirements, considerable  progress has been
made in performing radiative transfer simulations of  ionization maps
of representative regions of the Universe 
(see e.g. \citeNP{gnedin00,cfw03,pn05,impmsa06,ica++06,imsp07,mlz+07}).
Most radiative transfer simulations are, however, still  rather limited in
dynamic range and equally important also limited in their ability to
explore the large parameter space of plausible assumptions regarding
the spatial distribution and time evolution of the ionizing
emissivity. 

This is one of the reasons why modelling  the evolution of ionized
regions  analytically using excursion-set-like formalisms has become a
widely used and useful tool ({\it e.g.} \citeNP{fzh04b}). Such methods are well
adapted to  obtain estimates of  the size distribution of ionized regions for
arbitrary  models of the luminosity function and time evolution of the
ionizing emissivity. One has, however, to keep in mind that these
models make  quite drastic simplifying assumptions. The shapes of
ionized bubbles are e.g. assumed to be spherical and the (relative)
spatial distribution of sources and sinks of ionizing radiation are
not properly taken into account. 

Ideally,  one would like to compare realistic models of  the
ionization state  of the IGM  with a large dynamic range for a wide
range of assumptions  with future observations. For this purpose a
variety of  semi-numeric formalisms  have recently been proposed which
are based on performing an excursion-set formalism on the initial
Gaussian random field. The models  predict the spatial distribution of
the (integrated) ionizing emissivity as well as the spatial
distribution of  ionized
regions \cite{zlm+07,mf07,gw08}. They incorporate many of the relevant
physical  processes and allow the modeller to produce  21cm maps for
representative volumes of the Universe with a modest computational effort.

These studies suggest
that  reionization proceeds strictly 
inside-out with dense regions ionized
first and reionization slowly progressing into the large underdense
region as time goes on \cite{fzh04b,wm07,mf07,mlz+07}.  This  appears,
however,  in conflict with what is expected and observed for the
post-overlap phase where the low-density regions are found to be
highly ionized while high-density  regions remain neutral because of
their high recombination rate \cite{mhr00,wl03,cf05,cf06b}.  These
neutral regions determine the photon mean free path and manifest
themselves as Lyman-limit systems in QSO absorption spectra. Based on
the this low-redshift intuition derived from studying  the
intergalactic medium at $z\sim 2-4$ with Ly$\alpha$ forest data,  one
is thus  drawn to the conclusion that reionization  must have
proceeded -- at least to some extent -- outside-in rather than inside-out in the final
stages. The obvious suspect for resolving this apparent  contradiction
is the role recombinations play  in these simulations. Most of these models
assume a spatially uniform distribution  of recombinations and hence
do not take into account the self-shielding and shadowing of
high-density regions. \citeN{fo05}  have attempted to model this   by
introducing the concept of recombination-limited bubbles in analytic
studies of  the size distribution of ionized bubbles, which has been
implemented in simulations by \citeN{lzfmhz08}. As we will show
in this paper it is   important to realistically  model the spatial
distribution of the sinks of ionizing radiation due to recombinations 
when modelling the topology of reionization.  Many of the models also assume that
reionization occurs rather fast diminishing the relative importance of
recombinations. This appears,  however, to be in conflict with the
ionizing emissivity   inferred  from the opacity of the \lya\ forest 
in QSO absorption spectra which suggests that reionization occurs
slowly in a photon-starved regime \cite{bh07}.

We will study here the effects of the inhomogeneous spatial
distribution of recombinations on ionization maps and present a
consistent picture of reionization combining the concepts of growing
bubbles in the pre-overlap phase with the expected  presence of
neutral clumps  in the post-overlap phase. Our modelling  is similar
in spirit to  other semi-numerical  models of this kind  and in many
aspects we (need to) make similar  approximations  and
simplifications. 

In order to determine whether a high-density clump can remain neutral
or self-shielded against ionizing radiation, it is  necessary to
determine its position with respect to the nearest sources of ionizing
photons. An important requirement for a realistic model  of the
spatial distribution of density-dependent recombinations is thus  (i)
a realistic representation of the baryon distribution  and more
importantly, (ii) the location of the sources of ionizing sources
with respect to the density field. Note that we will here  concentrate on a
qualitative  understanding of the physical effects of a spatially
inhomogeneous distribution of recombinations on the
topology of reionization.

The paper is organized as follows: We describe our method for
generating the ionization maps in Section 2. 
Section 3 discusses our  main results  for the modelling of
a single source and representative volumes  of the Universe. In
Section 4 we check the consistency of our modelling with 
\lya\ forest data. In Section 5 we present predictions for the 
evolution of the  power spectrum and probability distribution of 
21cm emission and discuss prospects for the first generation
low-frequency instruments LOFAR and MWA. Section 6 contains our
conclusions. Throughout the paper,  we
assume a flat Universe with cosmological parameters  $\Omega_m =
0.26$, $\Omega_{\Lambda} = 0.74$, $\Omega_b h^2 = 0.022$, and
$h=0.73$.  The  parameters defining the linear dark  matter power
spectrum we use are $\sigma_8=0.9$, $ n_s=1$, $\de n_s/\de \ln k =0$
\cite{vhl06}.

\begin{figure*}
\rotatebox{270}{\resizebox{0.4\textwidth}{!}{\includegraphics{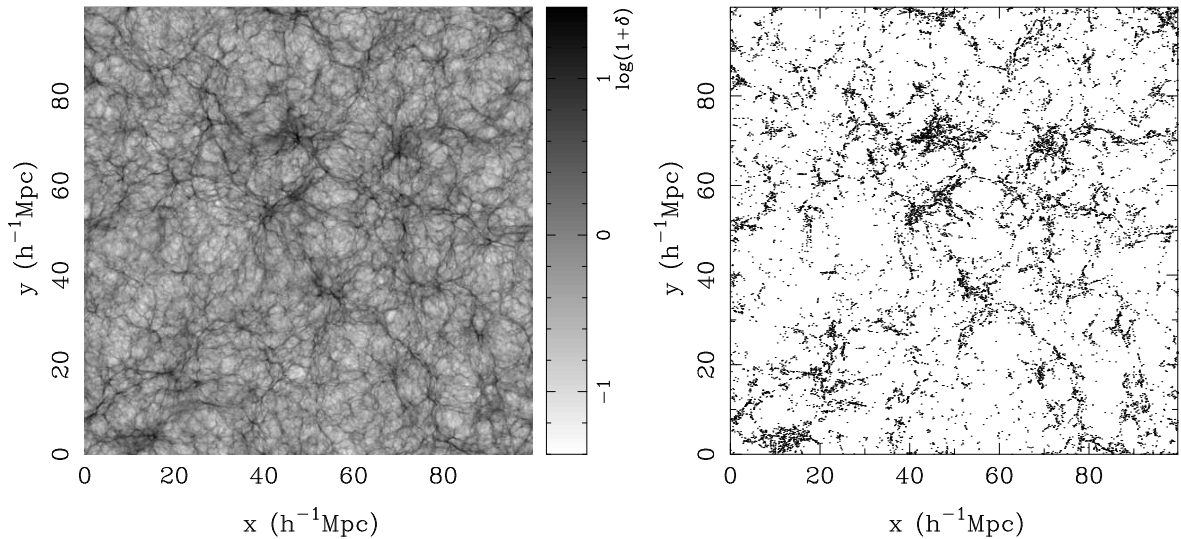}}}
\caption{The density field (left panel) and the location of collapsed
haloes  (right panel) at $z=6$ for our fiducial simulation (see text
for details).  The thickness of the slice shown is 1$h^{-1}$Mpc.}
\label{fig:denplushalo}
\end{figure*}

\section{Method}

Our  method of constructing ionization fields at a given redshift
consists of four steps:(i) generating the dark matter density field,
(ii) identifying the location and size of collapsed objects (haloes)
within the simulation box, (iii) assigning  photon luminosities to the
haloes and (iv) generating  maps of ionized regions from the spatial
distribution of the ionizing emissivity.  We discuss each of these
steps in the following subsections.

\subsection{Simulating the dark matter density field}

We obtain our representations of the dark matter density distribution
using the Zel'dovich approximation. We first generate an initial
linear density field (as is routinely done in N-body simulations) and
then displace the particles from their initial (Lagrangian)
coordinates ${\bf q}$  using the relation \be {\bf x}({\bf q},z) =
{\bf q} + D_+(z) {\bf \nabla_q} \phi({\bf q}), \e where $\phi({\bf
q})$ is the initial velocity potential and  $D_+(z)$ is the growth
factor of linear dark matter density perturbations.

The advantage of the Zel'dovich approximation is its  much larger speed
compared to  a typical N-body simulation of comparable size. This
allows us to produce ionization maps with a very large dynamic range
at a modest computational cost.  As we will show later (and has been
shown before)  the density field obtained in this way is  a
reasonable approximation to that obtained using full N-body
simulations, particularly at high redshifts. 

In  Figure \ref{fig:denplushalo} we show the projected two-dimensional
density field of a 1$h^{-1}$ Mpc thick slice through our fiducial
simulation.  One can clearly identify the expected filamentary
structures and voids, though  the range of overdensities achieved at
small scales are typically less than those obtained using full
simulations.  A possible objection against the use of the Zel'dovich
approximation  is that it becomes invalid once  shell-crossing
occurs. Note, however, that at the  redshifts and at scales of our
interest ($\gtrsim 1$Mpc) this occurs rarely.  A more detailed
comparison of the dark matter  distribution obtained  with the
Zel'dovich approximation with that of N-body simulations is performed
in Appendix \ref{app:compzeldo}.

Our fiducial simulation volume is a periodic box of length $100
h^{-1}$ Mpc  (comoving) containing $1000^3$ dark matter particles
which corresponds to a particle mass of $M_{\rm part} = 7.22 \times
10^7 h^{-1} M_{\odot}$. In order to check for numerical convergence,
we have  run further simulations with differing box sizes and particle
numbers; these are described and discussed in Appendix
\ref{app:numres}.

\subsection{Identifying  haloes}

\nocite{st02}
\begin{figure}
\rotatebox{270}{\resizebox{0.45\textwidth}{!}{\includegraphics{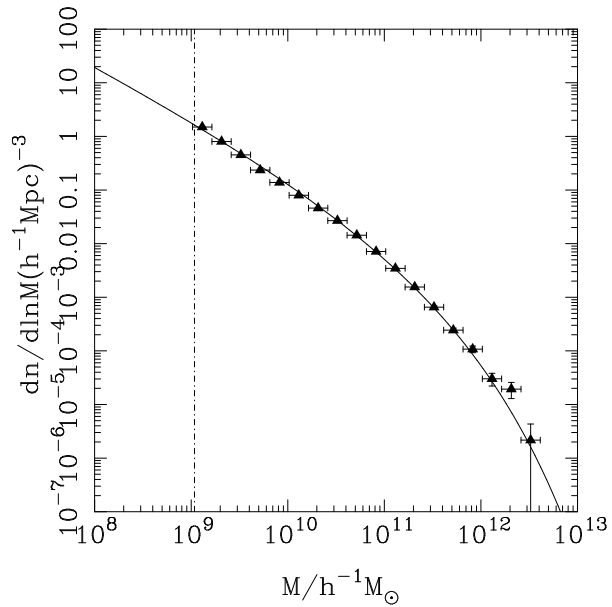}}}
\caption{The halo mass function at $z=6$. The points with errorbars
show the results from our simulation; the vertical errors correspond
to the statistical uncertainties while the horizontal errors denote
the bin size. The solid curve is the theoretical mass function of
Sheth \& Tormen (2002) with the  fitting function adopted from
Jenkins et al. (2001).}
\label{fig:comparehalomassfunc}
\end{figure}

The identification of haloes within the simulation box is performed
using a standard Friends-of-friends (FoF) algorithm \cite{defw85}.
Usually the mass function of haloes identified using the FoF algorithm
with a fixed linking length $b \approx 0.2$ (in units of mean
inter-particle separation) is found to give an excellent match to the
theoretical halo mass function for masses as small as $\sim 15-20
M_{\rm part}$ (for a recent example, see \citeNP{swj++05}).
Unfortunately, the use of the standard linking length  fails when
applied to  the density field generated using the Zel'dovich
approximation due to the more diffuse matter distribution in high
density regions.  However, if we use  the FoF algorithm with an
adaptive linking length which, depending on the redshift of interest,
lies  in the range  $0.30 \lesssim b \lesssim 0.37$ we get very
reasonable results.  Note, that the fact that the haloes  do not have
the correct density profile  is not a major concern here.  For our
purposes it  is sufficient to obtain the correct  location and mass of
the haloes with respect to the density field. 
Our method is a somewhat simpler version of algorithms based on 
Lagrangian perturbation theory \cite{mttgqs02,ss02}. The main difference
is that we identify haloes in  Eulerian
space using a FoF group-finder algorithm.

The location of haloes within a slice of  the simulation box is shown
in the right panel of Figure \ref{fig:denplushalo}. As expected  the
haloes trace the high-density peaks of the field.  The comoving number
density  of haloes per unit logarithmic mass $\de n/\de \ln M$  at
$z=6$ is shown in  Figure \ref{fig:comparehalomassfunc}  by the points
with error-bars. The theoretical mass function as predicted by
\citeN{jfw++01} is shown as the solid curve. The agreement  is
excellent over a  wide mass range $10^9 \lesssim M/(h^{-1} M_{\odot})
\lesssim 10^{13}$.  The lower mass limit corresponds to $\sim 10
M_{\rm part}$.

\subsection{Assigning ionizing luminosities}

Observationally little is known how the ionizing luminosity varies
with galaxy properties \cite{iid06,cpg07,gkc08}.  Models for
reionization thus often assume  that the ionizing  luminosity from
galaxies scales as the halo  mass with an efficiency  factor chosen
such that the integrated ionizing emissivity is  sufficient to
complete reionization.

We do the same and  and assume that the number of ionizing photons
contributed by a halo of mass $M$ is given by \be N_{\gamma}(M) =
N_{\rm ion} \f{M}{m_H},
\label{eq:ngamma_nion} \e where $m_H$ is the hydrogen mass and $N_{\rm
ion}$ is a dimensionless constant. The significance of $N_{\rm ion}$
can be understood by estimating the globally averaged comoving photon
density \be n_{\gamma} = \int_{M_{\rm min}}^{\infty} \de M \f{\de
n}{\de M} N_{\gamma}(M), \e which can be written in terms of the
fraction of mass in collapsed objects 
\be f_{\rm coll} = \rho_m^{-1}
\int_{M_{\rm min}}^{\infty} \de M \f{\de n}{\de M} M, \e  
as \be
n_{\gamma} = N_{\rm ion} \f{n_H}{1-Y_{\rm He}} ~ f_{\rm coll},
\label{eq:ngamma_fcoll} \e where $n_H$ is the comoving hydrogen
density. $N_{\rm ion}$ is the number of photons entering the IGM
per baryon in collapsed objects  \cite{wl07}.  It is 
determined by a combination of star-forming efficiency within the
halo, number of photons produced per unit stellar mass and the photon
escape fraction.  Note that the helium weight fraction $Y_{\rm He}$
could equally well be absorbed  into the definition of $N_{\rm ion}$; in that
case it would be equivalent to the parameter $\zeta$ used by
\citeN{fzh04b} and \citeN{mf07}.  The analysis presented in this paper
is applicable for any functional form of $N_{\gamma}(M)$. For example, one can
include QSOs in the analysis by simply assuming that they form in
haloes above a given mass $M_{\rm QSO}$, i.e., $N_{\gamma,{\rm QSO}}(M) =
N_{\rm ion, QSO}(M) \Theta(M_{\rm QSO}/M)$, where 
\bear \Theta(x) &=& 1,
~~~~\mbox{if $x < 1$},\nline &=& 0, ~~~~\mbox{otherwise}.
\label{eq:theta} \ear
and $N_{\rm ion, QSO}(M)$ is the number of ionizing photons produced within
a QSO-hosting halo of mass $M > M_{\rm QSO}$.

\subsection{Generating the ionization field}

Once the location and mass of haloes are known and the functional form 
of $N_{\gamma}(M)$ is assigned, the ionization field can be
generated using an excursion-set formalism as introduced by
\citeN{fzh04b}. First we  determine whether a given (spherical)
region is able to ``self-ionize''. We  estimate the mean number
density of photons  $\langle n_{\gamma}({\bf x}) \rangle_R$ within a
spherical region of radius $R$ around a point ${\bf x}$ and compare it 
with the corresponding spherically-averaged  hydrogen number density
$\langle n_H({\bf x}) \rangle_R$.  The condition for a point ${\bf x}$
to be ionized is that 
\be 
\langle n_{\gamma}({\bf x}) \rangle_R \geq
\langle n_H({\bf x}) \rangle_R (1 + \bar{N}_{\rm rec})
\label{eq:ngnh} 
\e 
for any $R$, where $\bar{N}_{\rm rec}$ is the
average number of recombinations per hydrogen atom in the IGM. For the
simple model where $N_{\gamma}(M) \propto M$, the above condition
translates to \be \langle f_{\rm coll}({\bf x}) \rangle_R \geq
\left(\f{N_{\rm ion}}{1-Y_{\rm He}}\right)^{-1} (1 + \bar{N}_{\rm
rec}),
\label{eq:fcoll_nrec} \e which is identical to what is used in
\citeN{mf07}.  Points which do not satisfy the above condition are
assigned a ionized fraction $Q_i({\bf x}) = \langle n_{\gamma}({\bf
x}) \rangle_{R_{\rm min}} / \langle n_H({\bf x}) \rangle_{R_{\rm
min}}$, where $R_{\rm min}$ is  the spatial resolution of the
simulation. This is important  to account for the HII regions
not resolved by the resolution of the simulations \cite{gw08}. Note
also that the effect of spatially uniform recombinations  (i.e., the
$1 + \bar{N}_{\rm rec}$ term) can be absorbed within the definition of
$N_{\rm ion}$.

Before identifying ionized regions,  we smooth the density
field to a grid-size of 1$h^{-1}$Mpc, corresponding to $100^3$ grid
points in the box. We do this in order  to smooth out the smaller
scales which are generally comparable to the largest halo sizes where
the  Zel'dovich approximation ceases to provide a good  approximation
for the evolution of the matter distribution.

The quantity $\langle n_{\gamma}({\bf x}) \rangle_R$ is estimated as
\be \langle n_{\gamma}({\bf x}) \rangle_R = \left(\f{4 \pi
R^3}{3}\right)^{-1} \sum_i N_{\gamma}(M_i) \Theta\left(\f{|{\bf x -
x_i}|}{R}\right),
\label{eq:nphot_R} \e where the sum is over all luminous haloes and
$\Theta$ is defined in equation (\ref{eq:theta}).  
$\langle n_{\gamma}({\bf x}) \rangle_R$  essentially
measures the contribution of ionizing photons at ${\bf x}$ arising from all the sources
within a radius $R$ around the point.  When  dealing with a small
number of sources, the summation in the above equation can be done
directly for every point in the simulation box.  When the number of
sources becomes large, direct summation is computationally expensive.
We therefore convert the point source distribution into a field. The
filtering is then done in  Fourier space.  The spherically-averaged
hydrogen number density $\langle n_H({\bf x}) \rangle_R$ is computed
by assuming that the hydrogen distribution follows the dark matter
distribution and then filtering the density field over a scale $R$
\cite{mf07}.

We should mention here that our method of obtaining the ionization
field follows that  of \citeN{mf07} with one notable difference.  For
a given $R$ and ${\bf x}$, we assume only the  pixel at the centre 
of the sphere with radius R to be
ionized when the threshold  (\ref{eq:ngnh})  is crossed while
\citeN{mf07} assume the entire filter sphere to be ionized. In this respect, our
modelling is   similar to that of \citeN{zlm+07}. We have
checked our method for isolated sources and found a good match with
theoretical expectations (to be discussed in \ref{sec:qsobubble}). In
the case where all (or most of) the  identified haloes contribute to
reionization, we find that the mass-averaged neutral fraction obtained
through our method agrees with the theoretical value  $1 - N_{\rm ion}
f_{\rm coll}/(1 - Y_{\rm He})$ to within 15 per cent.  This difference
arises because the semi-numeric schemes do not conserve
the number of photons within overlapping ionized regions
\cite{zlm+07}.

\subsection{Implementing a more realistic  inhomogeneous spatial distribution of
sinks of ionizing radiation due to recombinations}
\label{sec:recombination}

So far we have accounted for recombinations simply by multiplying  the
number of ionizing photons produced by a universal  factor $1 +
\bar{N}_{\rm rec}$  which does not depend on location. This corresponds
to assuming a homogeneous spatial  distribution of recombinations. In
reality the spatial distribution of sinks of ionizing radiation due to
recombinations will
be highly inhomogeneous. Even if a given spherical region contains
enough  photons to self-ionize, the high-density clumps within the
region will remain neutral for  a longer period because of their high
recombination rate and thus alter the nature of the  ionization
field. A  simple prescription to describe the presence of such neutral
clumps  by assuming that regions with overdensities above a critical
value ($\Delta > \Delta_i$)  remain neutral was  suggested by
\citeN{mhr00}. Unfortunately for our purpose this is also  not
appropriate  as many of the high-density regions are expected to harbour
ionizing sources.  Whether a region remains neutral will depend on
two competing factors, the local density (which determines the
recombination rate) and  the proximity to ionizing sources (which
determines the number of photons available). It is thus important to
include a realistic spatial distribution of  recombinations into the
formalisms for making ionization maps.

As a first approximation, one can incorporate recombinations within
the formalism by introducing a threshold condition similar to  equation
(\ref{eq:ngnh}), i.e.,
\be 
\langle \dot{n}_{\gamma}({\bf x}) \rangle_R \geq
{\cal C}_R \alpha_R \langle n_H({\bf x}) \rangle_R^2 ~ a^{-3}, 
\label{eq:ngrec} 
\e 
where $\dot{n}_{\gamma}({\bf x})$ is the comoving photon emissivity, 
$\alpha_R$ is the recombination rate at a temperature of $10^4$K.
Note that both the number densities $n_{\gamma}$ and $n_H$ are expressed
in comoving units.
The above condition, which is similar to that used by
\citeN{fo05} and \citeN{lzfmhz08}, expresses the fact that 
for a spherical region of radius $R$ to be ionized, one needs 
the ionizing photon emissivity to be
larger than the spherically-averaged recombination rate within the region.
The quantity ${\cal C}_R$ is the clumping factor which also 
takes into account the
fact that not all the points within the spherical region would contribute
to the recombination rate. For example, \citeN{fo05} consider that
high-density points with $\Delta > \Delta_i$
remain neutral and hence should not be counted
while computing the recombination rate within the region.
In that case ${\cal C}_R \propto \int_0^{\Delta_i} \de \Delta \Delta^2 
P_V(\Delta)$ is a measure of clumping factor provided by low-density
region only, where $P_V(\Delta)$ is the volume-weighted density
distribution of the IGM.

Another possible way of modelling the  recombinations in high-density
regions is to use a  self-shielding criterion.  In order to be
ionized, a given point should satisfy the condition that it cannot
remain self-shielded, i.e.,  
\be 
[n_{\rm HI}({\bf x}) a^{-3}] [L({\bf x}) a] \sigma_H \leq 1, 
\e 
where $n_{\rm HI}({\bf x})$ is the comoving number density of neutral
hydrogen at the given point, 
$L({\bf x})$ is the 
comoving size of the of the absorber and $\sigma_H$
is the  hydrogen photoionization cross section. 
In order to
estimate the  HI density for highly ionized regions,  we use the
photoionization equilibrium condition: $n_{\rm HI} = (\alpha_R/\Gamma)
n_H^2 a^{-3}$, where the photoionization rate is $\Gamma =
\dot{n}_{\gamma} a^{-3} \lambda_{\rm mfp} \sigma_H$.   

Estimating $\Gamma$ thus requires the knowledge of the emissivity 
$\dot{n}_{\gamma}$ and local mean free path
$\lambda_{\rm mfp}$. For a given filtering scale $R$, we equate it
to the mean free path, i.e., $\lambda_{\rm mfp} = R$
\cite{lzfmhz08}. Sources within a distance
$R$ then  contribute to the emissivity. If we assume that the 
fluctuations in the emissivity are  negligible for scales smaller than
the mean free path, we can write
\bear
\dot{n}_{\gamma}({\bf x}) &=& \left(\f{4 \pi
R^3}{3}\right)^{-1} \sum_i \dot{N}_{\gamma}(M_i) \Theta\left(\f{|{\bf x -
x_i}|}{R}\right)
\nline
& \equiv& \langle \dot{n}_{\gamma}({\bf x}) \rangle_R, 
\ear
where $\dot{N}_{\gamma}(M_i)$ is the photon production rate within
the halo with mass $M_i$ and the summation is over all haloes. The
photoionization rate is then given by
\be
\Gamma({\bf x}) = \langle \dot{n}_{\gamma}({\bf x}) 
\rangle_R a^{-3} R \sigma_H
\propto \sum_i \f{\dot{N}_{\gamma}(M_i)}{R^2} \Theta\left(\f{|{\bf x -
x_i}|}{R}\right).
\label{eq:GammaPI}
\e
The above equation expresses the fact 
that photons travel an average distance of $R$
from the source before being absorbed and the ionizing flux at the 
point of the absorber is diluted by a factor $R^{-2}$. Also 
implicit is the assumption that no photons are lost to recombination
within the region except those in the central cell which may lead to 
slight underestimate of the extent of self-shielded regions. On first
sight it may appear from the above equation that sources which are within distances
much shorter $R$ are not properly taken into account (the flux
from such sources would be less diluted than  implied by the
$R^{-2}$ factor).  However, one has to keep in mind that the procedure
is repeated for different values of $R$. Sources that are
closer to the point will thus  be taken into account for a smaller value
of $R$.

With the above approximations, one can write a new condition for a point
to be ionized, which is
\be 
\langle \dot{n}_{\gamma}({\bf x}) \rangle_R  \geq 
\f{\alpha_R n_H^2({\bf x})}{a^3} \f{L({\bf x})}{R}.  
\e 
There still remains the issue that the present formalism for
identifying ionized regions is based on the cumulative number of
photons $n_{\gamma}$, while balancing the recombination requires the
instantaneous rate of photon production $\dot{n}_{\gamma}$ (as seen 
in the previous equation). Any
detailed  model for the evolution of the ionizing emissivity  would
predict both these quantities self-consistently.  We would, however, 
like to incorporate recombinations here without  entering into the
complexities of the reionization history over a wide redshift range.
We thus integrate the above from the start of reionization so that 
the left hand side gives the integrated number of photons. The 
right hand side is significant only when recombinations are 
important and hence we can write the above relation in an
approximate way,
\be 
\langle n_{\gamma}({\bf x}) \rangle_R \geq 
n_H({\bf x}) \f{\epsilon t_H}{t_{\rm rec}({\bf x})} 
\f{L({\bf x})}{R}, 
\label{eq:ngnhtrec}
\e
where $t^{-1}_{\rm rec}({\bf x}) \equiv
\alpha_R n_H({\bf x}) a^{-3}$ is the local recombination timescale
and $\epsilon t_H$ is the timescale over which the recombination
term has significant
contribution with $t_H$ being the Hubble time.
Note that the parameter $\epsilon$,
which determines the time-scale over which recombinations are
significant, depends on the ionization and  thermal history at a
given location. We also still need to account for the effect of an
enhanced recombinations  due to  clumping on scales smaller than
resolved by our simulations. This is often done in the form 
of a  sub-grid clumping factor, which should be of the order
(but larger) than unity \cite{bh07} and   can be absorbed within
the unknown parameter $\epsilon$.  Short of doing the full radiative
transfer problem we have little handle for a rigorous estimate of
$\epsilon$. We will thus   take it to be independent of ${\bf x}$ and
study the results for a couple of values, namely, 0.5 and 1.0. A value
of $\epsilon=1.0$  implies that recombinations are
significant over a Hubble time. These values of 
$\epsilon$ where chosen in order to simulate a model 
in the ``photon-starved'' regime of reionization 
suggested by the \lya\ forest data (\citeNP{bh07}, section 4). 
Smaller values of $\epsilon$ should correspond to the 
more  rapid reionization implemented in many published numerical
simulations where recombinations are less important.
Note further that we have absorbed the sub-grid clumping factor within
$\epsilon$ which is typically larger than unity and thus
$\epsilon \sim 1.0$ may not be that unreasonable. 

The only parameter which remains to be discussed is 
the size of the absorber which determines
the neutral hydrogen column density. An obvious choice 
for this is the local Jeans length $L_J({\bf x})$ \cite{schaye01}, which
depends on the temperature and density.  We assume  here a uniform  temperature 
of  $10^4$K. The Jeans length scales then as  $L_J \propto \Delta^{-1/2}$. 
Any uncertainty in the value of the absorber size (e.g., those arising from the
geometry of the object or a different value of temperature) 
would again be absorbed within 
the unknown parameter $\epsilon$. The ionized cells are identified 
using the two threshold conditions (\ref{eq:ngnh}) and (\ref{eq:ngnhtrec});
we find that the barrier corresponding to (\ref{eq:ngrec}) is almost always
weaker than (\ref{eq:ngnhtrec}) and thus does not
make much difference to the results.

Finally, we would like to point out that, while comparing the 
number of available photons to the recombination rate at
a given point, one should exclude the collapsed 
gas residing within the halo.
However, this affects only a handful number of cells within the box. The 
reason is not difficult to understand -- for cells where the
collapsed fraction is higher than, say 5-10 per cent (depending on
the exact value of $N_{\rm ion}$ being used), 
the cell usually produces enough  
photons to ionize itself and also overcome the self-shielding 
criterion. In other words, cells with a collapsed mass fraction 
$> 5-10$ per cent  would anyway 
be flagged as ionized when we use a filtering scale $R$ of the
order of the cell size. For cells with a collapsed mass fraction lower than
this, it hardly makes any difference whether we include the halo gas into the  
recombination budget (changes of the order of a few per cent only).

Note that our modelling probably somewhat overestimates the size 
of individual self-shielded  regions. This should, however, at
least partially be compensated by the fact that recombinations 
will occur outside of self-shielded regions and that our
simulations lack the self-shielded regions expected to be hosted by 
DM haloes with masses below the resolution limit of our simulations.

\subsection{Other radiative transfer effects: shadowing}

\begin{figure*}
\rotatebox{270}{\resizebox{0.27\textwidth}{!}{\includegraphics{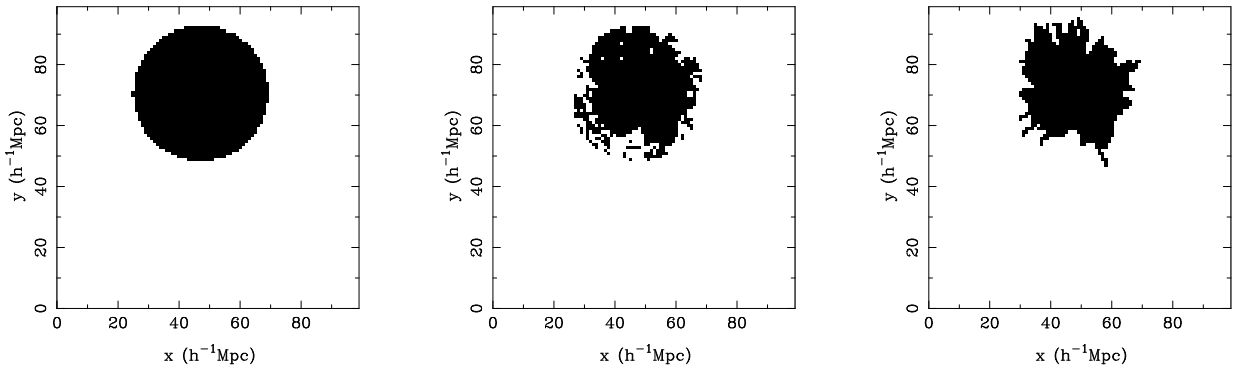}}}
\caption{Ionization maps for a single source (QSO)  with(middle
panel) and without (left panel) a spatially inhomogeneous
distribution of recombinations.  The right panel includes the additional effect of
shadowing.  The thickness of the slice shown is 1$h^{-1}$Mpc.}
\label{fig:plot_qso}
\end{figure*}

There are various radiative transfer effects which have not been taken
into account in our simplified treatment. The most important is  the
effect of ``shadowing''. High density clumps which are self-shielded
from ionizing photons will  not allow photon propagation to the other
side of the source. Such  shadowing effects can only be  incorporated
using some  form of ray-tracing algorithm which is beyond the scope of
the modelling here. We have studied the effect of shadowing  for a
single isolated source using a simple-minded ray-tracing algorithm;
the details are presented in Section \ref{sec:qsobubble}.

\subsection{Computational requirements}

The code used for this work has been  parallelized for shared-memory machines
using OpenMP. The simulations were run on COSMOS, a SGI Altix 4700
supercomputer. For our fiducial simulation box with $1000^3$
particles, we used 32 processors with a total RAM of 32 GB to
store the particle positions and velocities.  Generating the 
initial gaussian random field took about 10 minutes, and  obtaining
the position and velocity data for a
particular redshift using the Zel'dovich approximation took less than an
hour. A substantial amount of time was required to  identify the
position of collapsed haloes using the FoF halo finder.  For a single
value of the linking length, the halo finder  takes about 100 minutes to
run using 32 processors. However, since we are using an adaptive
linking length, the whole process takes much longer, about 14
hours. Thus, for a given redshift, generating  the density and
velocity fields alongwith the location of the haloes takes somewhere
around 17 hours. We should mention here that the FoF algorithm, which
takes most of the time, is easily parallelizable and scales well if 
a larger  number of processors is used.

The ionization fields were  generated with lower resolution.  
If we smooth the box to about $500^3$ grid points, the
process  takes about 40 minutes for a single set of
parameters. Since we probe a wide range of  parameter space, we
usually work with a smaller number of grid points, say,  $200^3$ or
$100^3$; generating ionization maps  takes then around a
minute to complete.

\section{The Effect of spatially inhomogeneous recombinations on the
topology of reionization}
\label{sec:results}

\begin{figure}
\rotatebox{270}{\resizebox{0.4\textwidth}{!}{\includegraphics{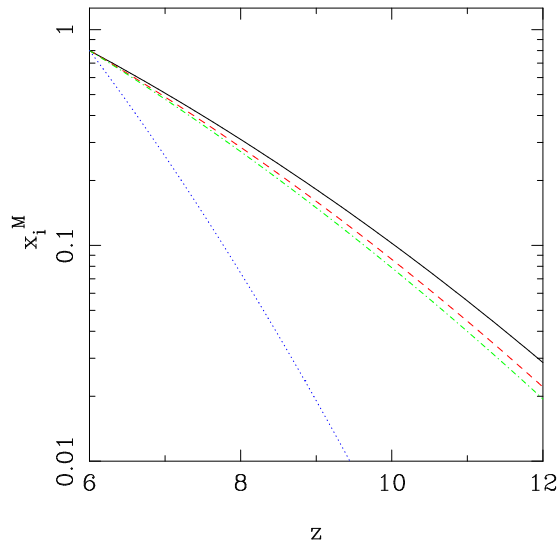}}}
\caption{Evolution of the mass-averaged ionized fraction for the
models with different assumptions regarding the spatial distribution
of sinks and sources of ionizing radiation as
described in section 3.2:  HR (solid curve), IR-0.5 (dashed curve),
IR-1.0 (dot-dashed curve) and IR-HM (dotted curve).  The photon
production efficiency in each model is normalised such that
$x_i^M(z=6) = 0.8$. }
\label{fig:xmevol}
\end{figure}

\begin{figure*}
\rotatebox{270}{\resizebox{0.95\textwidth}{!}{\includegraphics{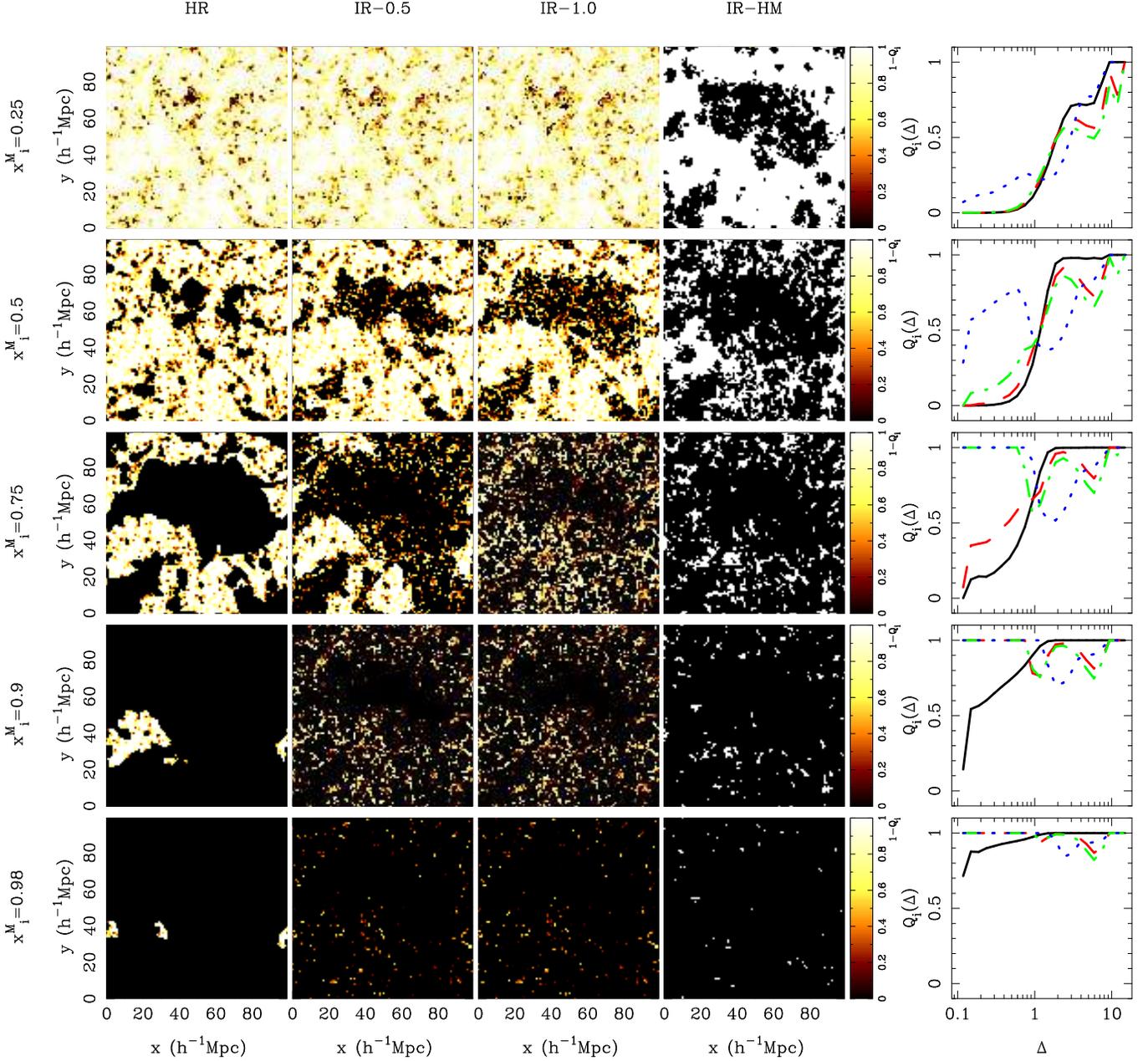}}}
\caption{Ionization maps  for a range of  mass-averaged ionized
fractions $x^M_i$ for the models with different assumptions regarding
the spatial distribution  of sinks and sources of ionizing radiation 
as described in section 3.2:   HR
(left-most panel), IR-0.5 (second panel), IR-1.0 (third panel) and
IR-HM (fourth panel).  The thickness of the slice shown is
1$h^{-1}$Mpc. The right-most panel shows the volume-averaged ionized
fraction $Q_i(\Delta)$ for the same models: HR (solid curve), IR-0.5
(dashed curve), IR-1.0 (dot-dashed curve) and IR-HM (dotted curve).}
\label{fig:plot_comparerec}
\end{figure*}

\begin{figure*}
\rotatebox{270}{\resizebox{0.3\textwidth}{!}{\includegraphics{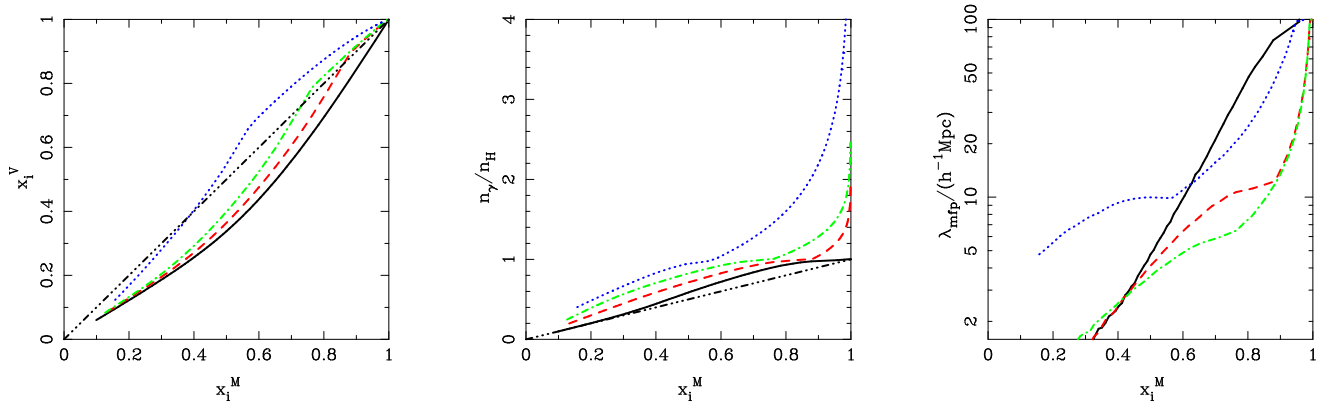}}}
\caption{Dependence of various quantities on the mass-averaged ionized
fraction $x_i^M$ for the different models: HR (solid curves), IR-0.5
(dashed curves), IR-1.0 (dot-dashed curves) and IR-HM (dotted
curves). The left panel shows the volume averaged ionized fraction
$x_i^V$. The dot-dot-dashed curve denotes  $x_i^V = x_i^M$.   The
middle panel shows the number of photons $n_{\gamma}$ produced per
hydrogen  atom $n_H$ in the IGM. The dot-dot-dashed curve denotes
$n_{\gamma}/n_H = x_i^M$.  The right panel shows the comoving mean
free path $\lambda_{\rm mfp}$ of ionizing  photons.}
\label{fig:plot_mfp}
\end{figure*}

\subsection{Test case: Stromgren sphere around a single source (QSO)}
\label{sec:qsobubble}

First, we consider the  case  of a single ionizing source with the
ionizing luminosity  of  a bright QSO in  the most massive halo ($M =
3.71 \times 10^{12} h^{-1} M_{\odot}$) in the simulation volume  as a
test case. The  ionizing  luminosity and the age of the QSO are chosen
such  that the ionized region has a comoving  radius of $\approx 22.5
h^{-1}$Mpc within an otherwise completely neutral and homogeneous IGM.
This corresponds to an ionized fraction of $x^M_i \approx 0.05$
averaged over the whole simulation volume. Ionization maps for
two-dimensional  slices centered on the ``QSO'' are shown in Figure
\ref{fig:plot_qso} for the case with and without an inhomogeneous
spatial distribution  of recombinations in the middle and left  panels
respectively. The right panel shows the same slice with the effects of
shadowing taken into account.

In the  left panel of the figure  where no (or only spatially
homogeneous recombinations have  been included) the ionized region is
spherical with  radius as expected for the assumed ionizing
luminosity. This confirms that our  method of generating ionization
fields is reasonably accurate for the case of spatially homogeneous
recombinations  and justifies our assumption that only the  central
pixel rather than the whole filtered sphere is ionized.

When the density dependence of  recombinations  are taken into account
allowing high-density  regions to stay neutral the appearance of the
ionized  region is very different due to the then very inhomogeneous
spatial distribution of sinks of ionizing radiation.   The resulting
ionized fraction also decreases  from  $x^M_i \approx 0.05$ to $x^M_i
\approx 0.03$,  despite the fact that the ionizing  luminosity of the
QSO  has the same value as before. This is simply due to the fact
that a larger  number of photons is needed to  overcome the
recombinations predominantly occurring in  high-density regions.  More
importantly, the shape of the ionized region is now  far from
spherical. The ionization fronts appear  to progress into  the
low-density regions while they are halted when high-density clumps are
encountered (see the left panel of Figure \ref{fig:denplushalo} for
the corresponding density field). However, we find that there are some
low-density pixels which lie in the shadow of a self-shielded clump
but are still ionized. This unphysical ``tunnelling'' of photons is a
limitation of  our  modelling  which does not take into account
shadowing effects.

The effect of shadowing is demonstrated  in  the right panel of Figure
\ref{fig:plot_qso}. In this case we have used a simple ray-tracing
algorithm where rays are going out from the source along all
directions. For each point along the ray, we check whether the local
photon density is sufficient  to ionize  hydrogen taking into account
recombinations, i.e., we check whether a point  can or cannot  be
self-shielded. The ray is terminated once it hits a self-shielding
pixel, thus forming a shadow on the other side of the high-density
point. The differences in the topology of the  resulting field are
obvious. The edges of the ionized bubble are more ragged when
shadowing is included. Note, however, that the difference in the
global ionized fraction is  only $\sim 0.0003$ or about  $\sim 1$ per
cent.  For representative volumes of the Universe the effect of
shadowing will be much less dramatic. Points  lying in the shadow of a
high-density  clump with respect to one ionizing source will generally
receive ionizing photons from sources in other directions.

\subsection{Modelling representative volumes of the Universe}
\label{sec:globalmaps}

We now discuss  ionization maps of representative volumes of the Universe
where  significant numbers of haloes (as opposed to a single
source)  host ionizing sources.  In order to investigate the effect of
a spatially  inhomogeneous distribution of sinks and sources of  ionizing
radiation and the speed
with which  reionization occurs  we consider four different models:
\begin{itemize}
\item{\it HR:} The spatial distribution of recombinations is assumed
to be homogeneous.  The condition for a region to be ionized is given
by equation (\ref{eq:ngamma_nion}), with $N_{\rm ion}$ being chosen so
as to give a defined  global mass-averaged ionized fraction $x^M_i$. 

\item{\it IR-0.5:} The spatial distribution of recombinations 
is assumed to be  inhomogeneous as  discussed in section 
\ref{sec:recombination}. 
The condition for a region  to be ionized is
determined by equation (\ref{eq:ngnh}) [which is 
same as in the HR model] and the self-shielding condition (\ref{eq:ngnhtrec}) 
with a value of $\epsilon=0.5$.
$N_{\rm ion}$ is adjusted to give  the same values of $x^M_i$ for all
four models.

\item{\it IR-1.0:} The same as the previous model but  with
$\epsilon=1.0$. The  effect of recombinations should be more prominent
than in the previous model.

\item{\it IR-HM:} The same as model IR-1.0 except that only  high mass
(HM) haloes with $M > 10^{11} h^{-1} M_{\odot}$ are ionizing sources.
This model  investigates the possibility  that ionizing photons within
lower mass haloes  may not be able to escape into the IGM efficiently
({\it e.g.} \citeNP{gnedin08}). These small haloes may still form stars,  
but in this model we assume that the ionizing photons are then absorbed 
within the interstellar medium and   
hence the galaxy remains mostly neutral, possibly contributing
significantly to the neutral hydrogen budget.
\end{itemize}

Before investigating the ionization maps generated using the above
models,  we first discuss  the predicted evolution of the global
mass-averaged ionized fraction $x_i^M$. To obtain the evolution of
$x_i^M$, we have calculated the collapsed mass fraction $f_{\rm coll}$
using the theoretical Sheth-Tormen mass function assuming a value of
$M_{\rm min}$ as set by our fiducial simulation box. This means that
the effects of various feedback processes on star-formation have been
ignored.  We have then estimated the value of $x_i^M$ from  $f_{\rm
coll}$ using the relation $x_i^M = N_{\rm ion} f_{\rm coll}/(1 - Y_{\rm
He})$ for the HR  model.  For the models with an inhomogeneous spatial
distribution of recombinations  the above relation was
modified to $x_i^M = N_{\rm ion} f_{\rm coll}/(1 - Y_{\rm He}) \times
(1 + \epsilon t_H/\langle t_{\rm rec} \rangle)$.  The values of
$x_i^M$ computed analytically in this way  differ from those  obtained
using the full simulations by up to 15 per cent, however, the basic
trends and other conclusions remain unaffected. The corresponding
evolution of $x_i^M$ for the  four models is shown in Figure
\ref{fig:xmevol}. The value of $N_{\rm
ion}$  is chosen in each case such that  $x_i^M = 0.8$ at $z=6$.

For models  HR, IR-0.5 and IR-1.0  (which have the same $f_{\rm
coll}$ at a given $z$),  the evolution of $x_i^M$ is nearly
identical.  The growth of $x_i^M$ is slightly  more  rapid in 
model IR-1.0 and slightly slower in model HR than in  model IR-0.5
, but the differences are small. For the same distribution of haloes,
reionization progresses ``faster''   as the spatially inhomogeneous
recombinations become more important.  At high-$z$,
the average recombination time  is shorter than the Hubble time. As a
result reionization is less  efficient at early epochs when the
spatially inhomogeneous recombinations  are  included.  As expected
the evolution of $x_i^M$  is drastically faster  in model IR-HM, where
only rare massive haloes host ionizing sources.  The collapsed
fraction in this model  is significantly smaller than in the other
models,  particularly at high redshifts, and hence reionization is
initially delayed.

We now discuss the nature of the ionization maps for the different
models.  Note that we have kept the
halo distribution fixed at that corresponding to $z=6$ and have varied the
luminosities to obtain different $x^M_i$ at the same redshift. In
reality, however, the variation in $x^M_i$ is due to the evolution of
the halo distribution with  redshift. We have here chosen to keep the halo
distribution fixed in order to focus  on the effect of the different
way we treat the spatial distribution of sinks of  ionizing radiation in the
different models.

The ionization fields for different $x^M_i$ are shown in Figure
\ref{fig:plot_comparerec} with the  left-most panel corresponding to
model HR.  The second, third and fourth panels describe  the three
models with a spatially inhomogeneous distribution of sinks of
ionizing radiation due to recombinations (IR-0.5, IR-1.0 and IR-HM, respectively).  The
right-most panel shows the volume-averaged ionized fraction
$Q_i(\Delta)$  as a function of overdensity $\Delta$. Including the
effects of a spatially  inhomogeneous distribution of recombinations   
distinctively  changes the  topology of ionized
regions at fixed ionized mass fraction. 

Let us first concentrate on the three columns of panels on the left of
Figure \ref{fig:plot_comparerec}  corresponding to models  HR, IR-0.5
and IR-1.0, respectively. In all three models the ionizing radiation
originates from the same dark matter haloes.  The models differ   only
in their treatment of recombinations.  To reach  the value of $x_i^M$,
one requires higher  values of $N_{\rm ion}$  in models IR-0.5 and
IR-1.0 than in model HR  as more photons are required to  overcome
recombination in high-density regions.   When the ionized mass
fraction is small ($x_i^M = 0.25$), the maps look very  similar.  At
this  stage  most of the ionizing photons are ionizing  the
high-density 
structures which host the photon sources.  At the  later
stages of  reionization ($x_i^M > 0.5$) ,  however, the topology of
the ionized regions  becomes very different in the three models. In
model HR the topology of the ionized  regions is  significantly
``smoother''  than in the other models. The  
high-density regions in models  IR-0.5 and IR-1.0  remain
neutral for longer and  hence a larger number of photons per hydrogen
atom is required to reach the  same $x_i^M$. Due to the larger number
of ionizing photons per  hydrogen atom  the ionizing photons are able
to reach   low-density regions far away from  sources of ionizing
radiation before the average ionized mass fraction becomes large.  The
ionization maps of  model HR  show   much larger coherently ionized
regions  while many  neutral (or partially neutral) clumps  are
embedded within  the ionized regions in the models with a spatially
inhomogeneous distribution of the sinks  of ionizing radiation
due to recombinations. 
In the very late stages of
reionization, models  IR-0.5 and IR-1.0  are nearly identical. 

The dependence of the ionization state on density is shown in the
right-most panel; the solid, dashed and dot-dashed curves correspond
to models HR, IR-0.5 and  IR-1.0, respectively.  As already mentioned,
early on ($x_i^M \lesssim 0.5$) the three models  are similar,
while they start to differ at later stages of reionization. Initially
the  topology can be described as ``inside-out''.  High density
regions  are ionized first.  However, in  model HR, the ionization of
the high-density regions is fully completed   before the ionization
fronts proceed into the underdense voids,  which are the last regions
to be ionized. In the HR model reionization proceeds ``inside-out''
all the way through  the reionization process.In models  IR-0.5 and IR-1.0,
on the other hand, the  ionization fronts are trapped by
high-density clumps and  they therefore proceed into low-density voids
leaving behind islands of neutral high-density gas. The topology is
now much more complex and cannot be classified  simply either as
``inside-out'' or ``outside-in''.  Underdense regions   ($\Delta < 1$)
are completely ionized by the time $x_i^M \sim 0.75$, which is
expected as the effect of recombination is negligible within the
low-density regions.  For regions with $\Delta \gtrsim 1$,
recombinations are important  and more  than one photon is required to
keep the region ionized. The ionized fraction $Q_i(\Delta)$
therefore decreases around $\Delta \sim 1$. Higher overdensities
$\Delta \sim 2$  are found close to the  filamentary structures in the
density field.  These regions harbour  small mass haloes, i.e,. the
relatively faint ionizing  sources which are able to overcome
recombinations to some extent and are responsible for an increase in
the  value of $Q_i(\Delta)$ around $\Delta \sim 2$. We have
verified this explicitly by computing the collapsed mass 
fraction within such cells. In even  higher
density regions, the number of photons required   to keep the region
ionized  becomes much larger than unity and  cannot be provided by the
fainter sources. Regions with  overdensities $\Delta \sim 6$ tend thus
to remain neutral.  The extremely high-density regions ($\Delta > 10$)
represent the   overlapping of filaments and harbour the  most
massive/luminous sources.  These regions are   able to overcome the
high recombination rates prevalent there and hence  can remain
ionized.

The ionization  maps of model   IR-HM  is very similar to those of
model IR-0.5 and IR-1.0.  The reversal to reionization progressing
more  ``outside-in'' occurs somewhat earlier, which is most obvious when
investigating the rightmost column showing the ionization state as a
function of density  $Q_i(\Delta)$ (dotted curves).  In model IR-HM
the ionizing sources reside in rare massive dark matter haloes.
A significant number of high-density regions of moderate mass
are devoid of any ionizing photon sources locally and  are able to
remain self-shielded from  ionizing photons. This   is very different
from model  IR-0.5 and IR-1.0 where almost all the  high-density
regions  host ionizing sources and hence  cannot remain completely
neutral. Also note that the behaviour of  $Q_i(\Delta)$ for
intermediate overdensities is somewhat different  from that  in models
IR-0.5 and IR-1.0. There is   no peak around $\Delta \sim 2$ in model
IR-HM. Recall that the peak in the  other models is due to the fainter
sources present within filamentary structures. These low-mass sources
are absent in model  IR-HM  and hence the corresponding peak in
$Q_i(\Delta)$ does not appear.

At this point, let us briefly compare our results with other
published results, particularly regarding the typical value of 
overdensities which can remain self-shielded. For example, \citeN{fo05} 
have shown, using modelling 
based on  \citeN{mhr00}, that an overdensity $\Delta$ at $z=6$
would be self-shielded
only if the local photoionization rate $\Gamma_{-12} < (\Delta/60)^{3/2}$,
where $\Gamma_{-12}$ is the  photoionization rate in units of $10^{-12}$
s$^{-1}$. We have explicitly verified whether this condition is satisfied
in  every self-shielded region by estimating $\Gamma_{-12}$ using 
equation (\ref{eq:GammaPI}). We find considerable 
fluctuations in the local value of $\Gamma_{-12}$ (which as expected 
decreases as
reionization progresses and the mean free path rises) and there do 
remain regions 
where $\Gamma_{-12}$ is much lower than what is required to overcome the
self-shielding. 
To give an explicit example, for the IR-1.0 scenario, 
we find that regions far away from sources 
have  $\Gamma_{-12}$ as low as 0.002 for $x_i^M = 0.95$ when the global mean
is $\approx 0.1$. The range of  values of $\Gamma_{-12}$ is typically
larger than that found by \citeN{md08}, which is probably due to the 
difference in the space density of ionizing sources (the smallest 
haloes in the simulations of \citeN{md08} have a mass of
$\sim 10^8 M_{\odot}$, 
while our mass threshold is $\sim 10^9 M_{\odot}$).

Having demonstrated  that reionization should not progress in a simple
``inside-out''  manner when the inhomogeneous distribution of  
recombinations  is   taken into account, we now
discuss  various other  quantities of interest for the
different models. The dependence of these quantities on $x_i^M$ is
shown in Figure \ref{fig:plot_mfp}.

The left panel shows the volume-averaged ionized fraction $x_i^V$.  In
the HR model  (solid curve) the ionized volume fraction does not
exceed the ionized mass fraction ($x_i^V \leq x_i^M$) for the whole range of
$x_i^M$ confirming that  ionization is biased towards high-density
regions. The models with a spatially inhomogeneous distribution of
recombinations have  $x_i^V \leq x_i^M$ in the  early stages
of reionization  (i.e., low values of $x_i^M$), while the trend
reverses later on.  This is in line with what we  discussed earlier,
i.e., reionization proceeds  ``inside-out'' at early stages while the
situation is more complex later.  As expected the reversal of trend
occurs earlier in the IR-HM model than in the IR-0.5 and IR-1.0
models.  The values of $x_i^V$ are  higher in the IR-HM model than in
the IR-1.0 model  for  given $x_i^M$. High density regions are,  on
average,  more neutral in model  IR-HM,  hence a larger volume has to
become ionized to reach  the same $x_i^M$. Note that for large ionized
mass fraction (say $\ge 0.95$) our models will increasingly 
underestimate the ionized volume fraction due to insufficient 
resolution.

The middle panel shows the number of ionizing photons per hydrogen
atom  $n_{\gamma}/n_H$ reaching the  IGM. The first point to be noted
is that $n_{\gamma}/n_H$ closely follows the ionized fraction $x_i^M$
in model HR.  Deviations arising  from 
a moderate violation of photon conservation of  our  algorithm for 
identifying ionized region are  $\lesssim 15$ per cent.
Obviously, the
ratio $n_{\gamma}/n_H$ is higher than $x_i^M$ for the other models
where   sinks of ionizing radiation  due to recombinations are included. Extra photons are
required to reach  the  same ionized mass fraction.  The other crucial
difference between model HR   and the other three models  is that for
large ionized mass fractions $n_{\gamma}/n_H$ flattens for model HR
while  it steepens when inhomogeneous recombinations are
included. In model HR  low-density voids are
the last regions to be ionized and hence the ionized volume
increases  without significant  further need for photons. The situation
is exactly opposite for the other cases where most of the photons are
being absorbed within high-density regions (acting as ``sinks'') and
hence no significant rise in $x_i^M$ is found even though the number
of photons used up increases  rapidly.

Finally, we plot the dependence of the mean free path $\lambda_{\rm
mfp}$ in the right panel. To calculate $\lambda_{\rm mfp}$, we first
randomly choose a ionized pixel and calculate the distance to a
neutral pixel along  a randomly chosen direction; this should denote
the local mean free path for the chosen point.  This Monte Carlo
procedure is repeated for a  large number of points. The global mean
free path is then estimated in  two different ways: (i) $\lambda_{\rm
mfp}$ is  estimated as the average of the different local mean free
paths and (ii) $\lambda_{\rm mfp}$ is estimated as the median of the
local mean free path distribution. In most cases, both  methods give
nearly identical estimates. The curves plotted in the figure are
obtained using the median [method (ii)].

The dependence  of $\lambda_{\rm mfp}$ on the ionized mass fraction is
most easily understood in the HR model (solid curve) where it is
determined by the characteristic size of ionized regions.
$\lambda_{\rm mfp}$ rises  with $x_i^M$  essentially featureless until
it flattens when $\lambda_{\rm mfp}$ approaches  the size of the
simulation box.  The trends for models IR-0.5 (dashed curve) and
IR-1.0 (dot-dashed curve)   are  similar to that in model HR in the
early stages of reionization ($x_i^M < 0.4$).  However,  as
reionization progresses, the mean free path  in models  IR-0.5 and IR-1.0
is smaller  than that in the  HR  model. High density clumps limit  the
propagation of ionizing photons in these models.  The  mean free path
in the models with spatially inhomogeneous recombination  is thus not
determined by the sizes of ionized regions when $x_i^M$ is large. It
depends instead on the spatial covering factor of  high-density peaks.  Note
that the mean free path  in the IR-HM model is larger than that in
the IR-0.5 and IR-1.0 models  for  given $x_i^M$.  This is consistent
with the fact in these models  a larger volume has to be ionized to
reach  the same ionized mass fraction.  Note  the ``break''  
in the evolution of $\lambda_{\rm mfp}$ for the models with
inhomogeneous recombinations. This break  broadly  defines the epoch
when the  mean free path starts to be limited  by high-density clumps
rather than  the size of ionized regions.
 
We should mention here that 
is likely that we have  overestimated the sizes of the
self-shielded absorbers because of the 
limited spatial resolution of our simulations 
This should lead to an  underestimate of 
$\lambda_{\rm mfp}$. The limited resolution 
will, however,  at the same time,  result in an underestimate 
of the space density of
self-shielded regions as well as of  recombination outside of 
self-shielded absorbers. This should in turn have lead to an 
overestimate of the mean free path. 
The two effects should thus partially cancel. 
We have examined the effect of resolution on $\lambda_{\rm mfp}$
in Appendix \ref{app:numres} and found that our results do not 
change when the resolution is improved by a factor of two. 
The absolute values of the mean free path shown in figure 
\ref{fig:plot_mfp} should nevertheless  be treated with some caution
but our finding that the mean free path will evolve more
slowly if  recombinations are  important should be robust.

\section{Consistency with Ly$\alpha$ forest data at $\lowercase{z}\sim 6$}
 
Current observational constraints on the epoch of reionization  are
still rather limited. Studies of the \lya\ forest  in QSO absorption
spectra  have taught us that reionization  probably ended at around $z
\sim 6$ (\citeNP{fnswbpr02,fhr++04,fsb++06}; {\it cf} \citeNP{brs07}). 
As discussed by \citeN{miralda03} and  \citeN{bh07}, the emissivity inferred 
from the \lya\ forest data
corresponds to at  most  a few photons per hydrogen atom per Hubble
time.  \citeN{bh07} thus coined the term ``photon-starved'' to
describe the regime in which reionization appears to occur.  \citeN{bh07}
measured  the emissivity of ionizing photons to be roughly
constant in comoving units in the redshift range $2<z <6$.  They
pointed out that because of the rather  low emissivity of ionizing
photons reionization of hydrogen  most likely started early and
extends over a wide redshift range.  This sits  well with the rather
large Thomson optical depth inferred from studies of the cosmic
microwave background \cite{sbd++07,dkn++08,cfg08}. 
Predictions of ionization maps  should
obviously be consistent with available data. Enforcing consistency 
with the \lya\ forest data shrinks the allowed parameter space
considerably and we  therefore discuss now how our modelling fairs in
this respect.

In Table 1 we summarize the mean-free path of ionizing photons
$\lambda_{\rm mfp}$ and the inferred photoionization rate $\Gamma$  in
our three models  for two values of the volume  fraction of ionized
regions (at $z=6$) and two different assumptions for when
reionization has started at $z_{\rm re}=15$ and $z_{\rm re}=\infty$, respectively.  
The value of the mean free path of ionising photons and  the volume
fraction of ionized regions $x_i^V$  at  $z=6$ are  observationally still
very uncertain. \citeN{bh07}  estimate the mean free path
to be $\la 40~{\rm Mpc}$ and infer a  photoionization rate
$\Gamma_{12} \la 0.19$.   The  volume fraction of ionized regions  has
been  estimated to be $\gtrsim 0.5$ at $z \gtrsim 6.3$ from  the
evolution of Ly$\alpha$ luminosity function \cite{ksm++06} and GRB
spectrum \cite{tkk++06}, while the constraints from QSO absorption
line measurements at $z \lesssim 6$ are quoted to give  $1 - x_i^V
\gtrsim 10^{-4}$ \cite{fsb++06}.

For our  models with a spatially inhomogeneous distribution of
recombinations the  mean free path is reasonably consistent
with the estimate of \citeN{bh07}  if the  volume  fraction of
ionized regions is large (95\%).  For models HR and IR-HM on the other
hand, the estimated mean free path is consistent with the values in Table 1
if the volume  fraction of ionized regions at $z=6$ is low  (50\%).

We have estimated the  photoionization rate (in units of $10^{-12}$
s$^{-1}$) inferred from the photon emission rate $\dot{n}_{\gamma}$
and $\lambda_{\rm mfp}$ using \cite{bh07},
\bear \Gamma_{-12} &\approx& 10^{-51.2}
\f{\dot{n}_{\gamma}}{{\rm s}^{-1} {\rm Mpc}^{-3}}
\left(\f{\alpha_s}{3}\right) \left(\f{\alpha_s+3}{6}\right)^{-1}
\nline &\times& \left(\f{\lambda_{\rm mfp}}{40 {\rm Mpc}}\right)
\left(\f{1+z}{7}\right)^2,
\label{eq:gamma12} \ear 
where $\alpha_s$ is the spectral index of the
ionizing background (which we assume to be 3 consistent with stellar
sources of sub-solar metallicity). The results are shown in the two right-most 
columns in table 1. Note again that there could be inaccuracies
of $\lesssim 15$ per cent arising from moderate violations of photon 
conservation of our algorithm.
For models IR-0.5 and IR-1.0 we find reasonable agreement with the
inferred photoionization rate for $x_i^V=0.95$. 
On the other hand, models HR and IR-HM generally tend to overpredict
the photoionization rate when the assumed ionized fraction is large.
For smaller values of the ionized mass fraction  ($x_i^V=0.5$), 
these models are found to be consistent with the data.

\begin{table}
\caption{Mean free path and inferred photoionization rate for
  different ionized mass fractions and different redshifts for the
start of reionization.}
\begin{tabular}{|c|c|c|c|c|c|} 
\hline 
Model  & $x_i^V$ & $\lambda_{\rm mfp}$  & $n_{\gamma}/n_H$ &
\multicolumn{2}{|c|}{$\Gamma_{-12}$}\\
& &$h^{-1} {\rm Mpc}$& &$z_{\rm re}=\infty$ &$z_{\rm re}=15$\\ 
\hline 
HR     & 0.5     & 15     & 0.79   &  0.031   &  0.044 \\
       & 0.95    & 97     & 0.96   &  0.257   &  0.362 \\ 
\hline 
IR-0.5 & 0.5     & 8      & 0.82   &  0.016   &  0.023 \\ 
       & 0.95    & 21     & 1.14   &  0.063   &  0.089 \\ 
\hline
IR-1.0 & 0.5     & 5      & 0.89   &  0.011   &  0.015 \\ 
       & 0.95    & 21     & 1.40   &  0.076   &  0.107 \\ 
\hline 
IR-HM  & 0.5     & 12     & 0.91   &  0.026   &  0.036 \\ 
       & 0.95    & 56     & 2.20   &  0.330   &  0.465 \\ 
\hline
\end{tabular}
\label{tab:gamma12}
\end{table}

\section{Predictions for 21cm observations}

\subsection{The effect of the spatial distribution of sinks and the
luminosity of sources  on the 21cm power spectrum.}

\begin{figure*}
\rotatebox{270}{\resizebox{0.29\textwidth}{!}{\includegraphics{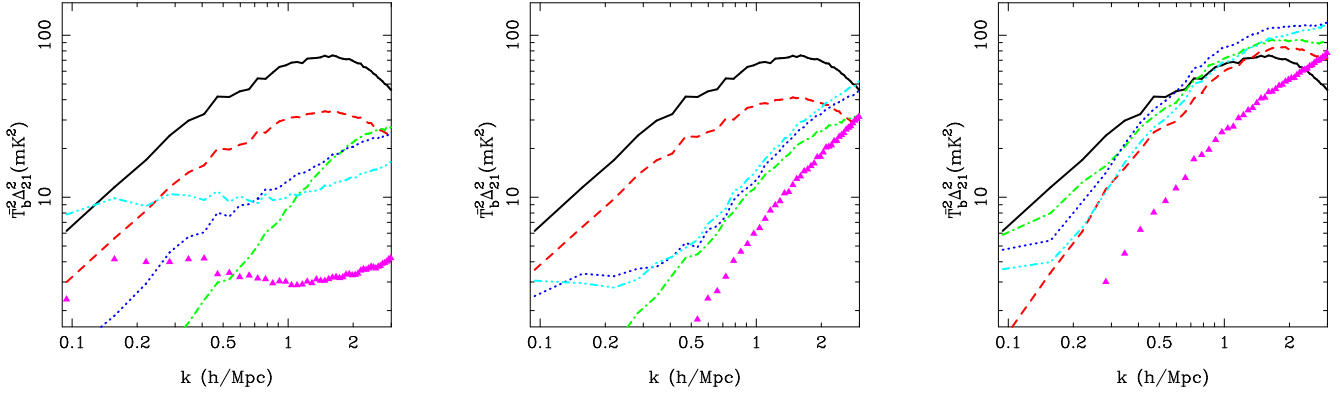}}}
\caption{The power spectrum of 21cm brightness temperature
fluctuations.  The panels from left to right show  models HR, IR
and IR-HM, respectively. In each panel, results are shown for
mass-averaged ionization fraction $x_i^M = $ 0 (solid), 0.1 (dashed),
0.3 (dot-dashed), 0.5 (dotted), 0.7 (dot-dot-dot-dashed) and 0.9
(triangles),  respectively.}
\label{fig:qipower}
\end{figure*}

\begin{figure*}
\rotatebox{270}{\resizebox{0.29\textwidth}{!}{\includegraphics{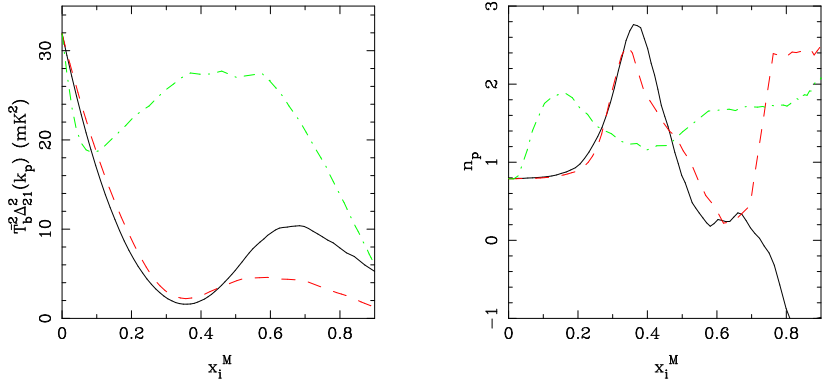}}}
\caption{The amplitude (left panel) and slope (right panel)  of the
power spectrum of 21cm brightness temperature fluctuations  as a
function of the mass-averaged ionized fraction  $x_i^M$ at wavenumber
$k = k_p \equiv 0.4 h {\rm Mpc}^{-1}$.  The solid, dashed and
dot-dashed curves represent models HR, IR and IR-HM, respectively.}
\label{fig:plotqipower}
\end{figure*}

\begin{figure*}
\rotatebox{270}{\resizebox{0.29\textwidth}{!}{\includegraphics{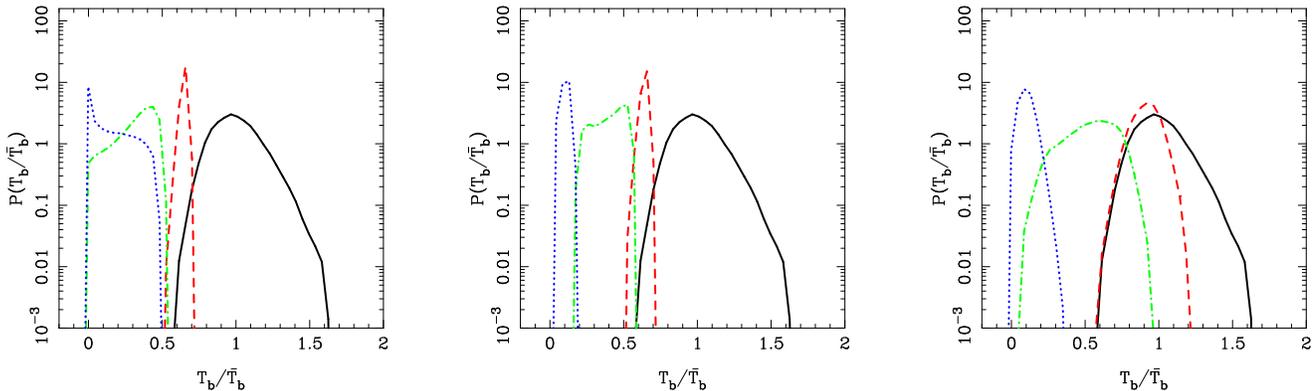}}}
\caption{The probability distribution of the dimensionless 21cm
brightness temperature fluctuations.  {\it Left panel:} Results
for model HR . The curves with peaks from right to left are for
mass averaged ionization fractions of  $x_i^M = 0, 0.35, 0.7, 0.9$
respectively.  {\it Middle panel:} Results for models  IR-1.0.
The curves with peaks from right to left represent cases
$x_i^M = 0, 0.35, 0.6, 0.9$ respectively.  {\it Right panel:} 
Results for model IR-HM. The curves with peaks from right to
left are for  $x_i^M = 0, 0.1, 0.45, 0.9$, respectively.  }
\label{fig:qipdf_smooth}
\end{figure*}

We have seen in Section \ref{sec:globalmaps}  that the models with
different assumptions regarding the spatial distribution of sinks and
sources of ionizing radiation  predict rather different topologies for
the  neutral hydrogen distribution, particularly in the  late stages
of reionization (with the exception  that models IR-0.5 and IR-1.0 are
nearly identical for $x_i^M > 0.9$). We now discuss the prospects of
investigating the effects of the spatial distribution of the sinks
and sources of ionizing radiation  with future low-frequency radio
observations of the redshifted 21cm line.  Since model IR-0.5 is
qualitatively very similar to model IR-1.0, we shall not discuss it
separately in this section.

The 21cm brightness temperature at a given location  ${\bf x}$
relative to the CMB can be approximated  as  
\be T_b({\bf x}) = \bar{T}_b x_{\rm HI}({\bf x}) \Delta({\bf x}), \e 
where we have assumed
that the spin temperature of hydrogen  is  much larger than the CMB
temperature. This  should be  a reasonable assumption once a significant
fraction (a few percent) of the volume/mass has  been ionized 
(\citeNP{sr90,tmmr00,cm03,bl05,sethi05,furlanetto06,fob06}; see \citeNP{pl08}
for an extensive recent
discussion  of the expected evolution of the spin temperature).   We
have also  ignored peculiar velocity effects which are small at
the scales relevant here (see, e.g., \citeNP{mf07}). 
The quantity $\bar{T}_b \approx 22 {\rm mK} [(1+z)/7]^{1/2}$ denotes
the brightness temperature for neutral gas at mean density.  By
definition, $\langle T_b({\bf x}) \rangle = \bar{T}_b (1 - x_i^M)$.

The first quantity of interest is  the power spectrum of  temperature
fluctuations which we define as  $\bar{T}_b^2 \Delta^2_{21}(k) \equiv
k^3 \langle T_b^2(k) \rangle/2 \pi^2$. The power spectrum  is plotted in Figure
\ref{fig:qipower} for our  models  for a range of values of $x_i^M$
(e.g. \citeNP{fzh04}). 
The panels from left to right show the power spectrum for models   HR, IR-1.0 and
IR-HM,  respectively. In each panel the power spectrum is  shown for
mass-averaged ionization fraction $x_i^M = $ 0 (solid), 0.1 (dashed),
0.3 (dot-dashed), 0.5 (dotted), 0.7 (dot-dot-dot-dashed) and 0.9
(triangles) respectively. For  $x_i^M = 0$, the brightness temperature
simply traces the DM fluctuations.

In model  HR  (left panel), the amplitude of the power spectrum
decreases from its initial value  until about $x_i^M = 0.3$
(dot-dashed curve).  The decrease of the fluctuation amplitude,
particularly at large scales,  occurs as  regions of high density
become ionized.  The decrease in amplitude is accompanied by a
steepening in slope,  consistent with the findings of
\citeN{lzmzh07}. It follows then  a reversal in trend.   The amplitude
rises  (particularly at large scales $k < 1.5 h$ Mpc$^{-1}$) and  the
slope becomes shallower. This is particularly evident  if the power
spectra for    $x_i^M = 0.3$  (dot-dashed curve)  and  $x_i^M = 0.5$
(dotted curve) are compared.  This is the phase when the the ionizing
radiation from collapsed objects ionizes the surrounding high-density
regions. The growth of ionized regions  boosts  the large scale power
and  flattens the slope of $\Delta_{21}^2(k)$.  The flattening of the
slope continues (and the  power spectrum becomes practically flat) as
the IGM becomes  more ionized while the amplitude  decreases at high
values of $x_i^M$.  In  model  HR there is nearly  equal power at all
scales in the late  stages of reionization and the fluctuation
amplitude  decreases as the neutral hydrogen content in the IGM
decreases.

In the early stages of reionization  ($x_i^M < 0.5$)  the evolution of
the 21cm power spectrum  in model  IR-1.0  (middle panel) is very
similar to that in model HR.  The similarity, however,  disappears
for  $x_i^M \geq 0.5$ when  the slope of $\Delta_{21}^2(k)$  steepens
rather than flattens.  The high-density neutral regions embedded
within the ionized regions  are responsible for a  considerable amount
of small-scale power in the 21cm power spectrum. At the same time, the
clumps limit the size of  coherently  ionized regions, thus keeping
the large-scale power low.  This pattern holds until the very end of
reionization.   The steepening of the slope in the later stages of
reionization in the models with an inhomogeneous spatial distribution
of recombinations is a signature of the more complex
topology which we had described in section 3.2. In the late stages
reionization proceeds much more ``outside-in'' than in model HR  and
this is clearly recognizable in  the 21cm power spectra.

A similar but more pronounced steepening of the slope of the 21cm
power spectrum  occurs in model IR-HM, where the emission of ionizing
photons is restricted to massive haloes.  Here reionization starts to
proceed  in a more ``outside-in'' fashion much earlier. The behaviour
of the amplitude  21cm spectrum at large scales ($k < 1 h {\rm
Mpc}^{-1}$) is less complicated than in the  other two models; there
is a slight dip in the power spectrum around $x_i^M = 0.1$ due to the
ionization of the high-density regions harbouring the sources of
ionizing radiation.  Otherwise the power spectrum  evolves very little
until  $x_i^M \sim 0.5$ and then the amplitude decreases with decreasing neutral
fraction. In model IR-HM the amplitude of the 21cm power spectrum is
generally somewhat higher than in the other two models.  This is due
to  reionization being driven by relatively highly clustered sources
in this model.

\subsection{ Evolution of slope and amplitude of the 21cm power
spectrum at scales probed  by LOFAR and MWA}

In the last section we developed a feeling for how the spatial
distribution of sinks and sources of ionizing radiation influence the
21cm power spectrum. We now discuss in more detail the possibility to
differentiate  observationally between different models with first
generation 21cm experiments  like LOFAR and MWA.   
The typical scales
probed by these experiments correspond to wavenumbers   $0.1 < k/{\rm
Mpc}^{-1} < 1$. Foreground subtraction will be a serious problem 
and it is not clear yet to how small and large scales it will be 
possible  to determine the power spectrum with reasonable accuracy. 
We follow \citeN{lzmzh07} and assume that the  optimum scale for studying 
21cm fluctuations with these instruments correspond to  $k \sim 0.3-0.5 {\rm
Mpc}^{-1}$ and use  a pivot scale of $k = k_p \equiv 0.4 h {\rm
Mpc}^{-1}$ in the following  to be definite.
This scale is well suited for a discrimination between
our  models.

The amplitude of the power spectrum at $k_p$ and its  slope $n_p
\equiv \de \ln \Delta_{21}^2(k_p)/\de \ln k$  as a function of the
ionized mass fraction $x_i^M$  are shown in the left and right panels
of Figure \ref{fig:plotqipower}. The evolution of  the different models
reflects our discussion in the last section.  The
evolution of the amplitude $\Delta_{21}^2(k_p)$  can be divided into
three phases. An initial  decrease in amplitude due to the early
ionization of high-density regions   (\citeNP{wm07}),  is followed by a
rise corresponding to a growth in patchiness and a final  fall due to
the elimination of neutral hydrogen.  In model HR-IM  (dot-dashed
curve) where  reionization is driven by rarer sources   the 21cm power
spectrum has significantly (a factor two or more) power than in the
other two models. The amplitude of the power spectrum and its
evolution at our pivot scale  contains valuable information about  the
spatial distribution of ionizing sources.

The amplitude of the power spectrum at our pivot scale appears,
however,  not to be a good indicator of  the spatial distribution of
the sinks of ionizing radiation. Models  HR and IR-1.0 models are  very
similar in this respect.  As discussed earlier the spatial
distribution of sinks has instead  a strong influence on the slope of
the power spectrum $n_p$ (right panel).
The evolution of the slope $n_p$ can be again divided into three
phases. An initial rise  due to the ionization of high-density regions
followed by a fall corresponding to the growth of patchiness. In the
third phase at $x_i^M > 0.65$ $n_p$ keeps on decreasing rapidly in
the HR model while it increases instead in  the models with a
spatially inhomogeneous distribution of sinks of ionizing radiation 
due to recombinations. 

By measuring the power spectrum and its slope at large scales ($k \sim
0.3-0.5 {\rm Mpc}^{-1}$) it should thus be possible to characterize
both the spatial distribution of sources and sinks of ionizing
radiation.  The detectability of the 21cm fluctuations obviously
depends  on the instrument noise and the ability to subtract
foreground emission.  Assuming a perfect removal of foreground
emission \citeN{mzzhf06} find typical values of detector noise  for
LOFAR and MWA at  $k \sim 0.4 h {\rm Mpc}^{-1}$ of  $\lesssim 5$
mK$^2$ at $z=6$ for  1000 hrs of observation. With such noise levels,
the power spectra for the HR and the IR-HM models should be detectable
with  reasonable confidence in the range $0.5 < x_i^M < 0.9$ and
$x_i^M < 0.9$ ,respectively.  The fluctuation amplitude in  model
IR-1.0 is  lower than in the other  two models discussed in this
section. At their peak value around $x_i^M \sim 0.6$ 21cm fluctuations
should nevertheless be detectable by LOFAR and MWA even for this
model.  Note that the values quoted here should be  only taken as
indicative.   Both the noise properties and the fluctuation amplitude
depend on redshift  For example in our models an ionized mass fraction
of $x_i^M \sim 0.6$ is  reached around $z \sim 8$ while our estimations
were performed  assuming  $z \sim 6$.

\subsection{The PDF of the 21cm brightness distribution}

We now briefly discuss the probability distribution  $P(T_b/\bar{T}_b)
\equiv P(\Delta~x_{\rm HI})$ of the dimensionless  brightness
temperature \cite{fzh04b}. In order to compute the distribution, we smooth the
brightness temperature $T_b$ over scales of $10 h^{-1}$ Mpc, as is
appropriate for the first generation 21cm experiments. The results are
shown in Figure \ref{fig:qipdf_smooth}.  The left panel shows the 21cm
PDF  for model  HR. The curves with peaks from right to left are for
$x_i^M = 0, 0.35, 0.7, 0.9$, respectively.  We have chosen the values
of $x_i^M$ such that they represent the characteristic points in the
evolution of the power spectrum  at large scales. The curve for $x_i^M = 0$
(solid)  obviously represents the dark matter PDF.  For
$x_i^M = 0.35$ (dashed curve),  the PDF has become significantly
narrower. This is again due to  the ionization of high-density regions
and corresponds to a  low-amplitude of the power spectrum. The
evolution of the PDF in model HR is consistent with the analytical
models of \citeN{wm07}. The PDF widens subsequently with increasing
$x_i^M$ as more regions are being ionized.  The behaviour is  similar
in model  IR-1.0  (middle panel) where  the curves with  peaks from
right to left represent  $x_i^M = 0, 0.35, 0.6, 0.9$, respectively.
The only difference  is a somewhat narrower  width of the distribution
than in model  HR  in the final stages of reionization ($x_i^M \sim
0.9$).  This is consistent with what is expected from  the evolution
of the 21cm power spectra at large scales.  The results for model
IR-HM are  shown in the right panel.  The curves with peaks from right
to left represent $x_i^M = 0, 0.1, 0.45, 0.9$, respectively.  As
expected, the PDF in this model is rather  different from that in the
other two models.  The PDF in model IR-HM has a wider distribution
compared to the other two models. The model predicts  $T_b/\bar{T}_b
\gtrsim 0.5$ even when the IGM is 50 per cent ionized by
mass. Unfortunately, it is not clear whether the first generation 21
cm experiments will  have enough sensitivity to constrain the shape of
the PDF.

\subsection{Comparison with other work} 

As discussed in the
introduction, there has been a number of recent studies which aim at
predicting the  21cm brightness distribution. These studies range
from radiative transfer simulation generally performed by
post-processing the density field of DM simulations 
\cite{cfw03,impmsa06,mips06,imsp07,mzzhf06,mlz+07,zlm+07} to semi-numerical
simulations \cite{mf07,aa07,gw08} similar in spirit to the work
presented here.  Most of these studies appear to agree that 
reionization occurs  inside-out all the way from the start 
until nearly the completion of reionization.  
\citeN{zlm+07}  have thereby shown that results for
semi-numeric schemes  based on collapsed mass fractions and variants
of the excursion set formalism to identify  regions which can
self-ionize  give  very similar results  to full radiative transfer
simulations if similar assumptions regarding the sources of ionizing
radiation are made. Most similar to our work here is probably the
work of \citeN{mlz+07} who have studied a wide  range of 
assumptions regarding the sources and sinks of ionizing radiation. 
When modelling the effects of sinks of ionizing radiation \citeN{mlz+07}
mainly  study mini-haloes, dark matter  haloes with potential
wells shallow enough so that they can be photo-evaporated by ionizing
photons. For these mini-haloes they find a noticeable but rather
small effect (see \citeNP{bh07} for a brief discussion of the
role of mini-haloes during reionization in the photon-starved regime). 
\citeN{mlz+07}, however, do not try to model  recombinations in 
high-density regions in deeper potential wells which can hold on to photo-ionized
gas in a way so that their  models  are likely to be 
consistent  with the \lya\ forest data. 
They generally find that sinks of
ionizing radiation and
their spatial distribution have little effect on the  topology of
reionization and the power spectrum. This is obviously quite
different  from our findings. There is a number of differences 
to our modelling but the most likely reason appears to be 
the following. The emissivity used in the models of \citeN{mlz+07} is 
rather high and reionization  proceeds quickly. This  
strongly diminishes the importance of recombination compared to our 
modelling of reionization in  the  photon-starved regime suggested 
by the Ly$\alpha$ forest data.

\section{Conclusions}

We have used here semi-numerical simulations to investigate  the role
of the  spatial distribution of sinks and sources of ionizing
radiation on the topology of hydrogen  reionization.  Our main results
are the following.

\begin{itemize}

\item The combination of Zel'dovich approximation, halo-finder and
excursion set formalism  is a powerful tool to calculate realistic
ionization maps with high dynamic range  at a  very  moderate
computational   cost. 

\item  Enforcing  consistency with the \lya\ forest data helps to
significantly shrink  the otherwise rather unconstrained
parameter space of models of reionization.  In the  photon-starved
regime of reionization suggested by the \lya\ forest data
recombinations  are much more important than in models with high
ionizing emissivity where
reionization occurs quickly.  Taking into account a realistic
spatially inhomogeneous distribution of sinks  of ionizing radiation
has a large effect on the topology of  reionization in the
photon-starved regime.

\item  Initially reionization proceeds inside-out with the
high-density regions hosting  the sources of ionizing sources becoming
ionized first. In the later stages of  photon-starved reionization the
sinks of ionizing region in our models remain neutral and  reionization proceeds
deep into the underdense regions before slowly  evaporating  denser
regions not hosting ionizing sources where recombinations are
important.  This  reversal to a more
outside-in progression in the late stages of reionization  is more
pronounced if the  emission of ionizing radiation is  restricted to
massive  highly-clustered and rare sources.
           
\item  If the emission of ionizing radiation is restricted to rare
sources   reionization proceeds more quickly and  the sizes  of
coherently ionized regions are  significantly larger.  The latter
results in an about factor two or more larger  mean free path for
ionizing photons. 

\item Like other studies we find that the amplitude of the 21cm power
spectrum and its  evolution in the later stages of reionization is
mainly sensitive to the space density of ionizing  sources.   The
sensitivity to the space density of ionizing sources is, however,  significantly
increased if a realistic spatially  inhomogeneous distribution of
sinks of ionizing radiation is taken  into           account. The slope of
the power  spectrum is very sensitive to the spatial distribution  of
sinks of ionizing radiation. 
           
\item  Measurements of the amplitude and slope of the 21cm power
spectrum at scales  corresponding  to    $k \sim  0.3 - 0.5 h{\rm
Mpc}^{-1}$  with the upcoming low-frequency   instruments LOFAR and
MWA have excellent prospects to reveal important
information on the spatial distribution of sinks and sources  of
ionizing radiation and the speed of  reionization if the daunting tasks
of accurate calibration  and foreground removal are mastered  successfully.
The PDF of the  21cm brightness distribution contains important
complimentary information. Measuring the PDF  will, however,
unfortunately most likely require higher sensitivity than can be
achieved with first generation 21cm experiments.                   
            
\end{itemize}

Our modelling here has involved  a number of significant
simplifications. The spatial  distribution of dark matter   modelled
in the Zel'dovich approximation was used as an proxy for the spatial
distribution  of the IGM. The ionizing emissivity  of sources and
recombination in dense region was modelled  only in an approximate
integrated fashion and  the dynamical effects of the ionization
radiation  on the gas were neglected. Despite the large particle
number used in the simulations resulting in a substantial dynamic
range there were still clear deficiencies in modelling high-density
regions and low-mass collapsed objects/mini-haloes.  
We nevertheless
think that our simulations  have caught the essential properties of
the topology of  the epoch of reionization. 
Our simulations suggest
that the idea that reionization proceeds strictly inside-out from 
beginning to nearly to the end may need revision if reionization
indeed occurs in a photon-starved regime as suggested by  the 
\lya\ forest data.

\section*{Acknowledgments}

We thank Tom Abel, Benedetta Ciardi,  Nick Gnedin, Ilian Iliev, Adam
Lidz, Avi Loeb, Matthew McQuinn, Jordi Miralda-Escud\'e and Paul 
Shapiro for valuable comments made at the 2008 Harvard
conference on 21cm Cosmology where part of this work was presented.  
This research was conducted in cooperation with SGI/Intel utilizing
the Altix 4800 supercomputer COSMOS at the Department of Applied
Mathematics and Theoretical Physics in Cambridge. COSMOS is a UK-CCC
facility which is supported by HEFCE and STFC/PPARC.  Part of the
simulations where performed  on the  Cambridge High Performance
Computing Cluster Darwin.

\bibliography{mnrasmnemonic,astropap-mod,reionization}
\bibliographystyle{mnras}

\appendix
\section{Comparison of different methods of generating the halo field}
\label{app:compzeldo}

\begin{figure*}
\rotatebox{270}{\resizebox{0.9\textwidth}{!}{\includegraphics{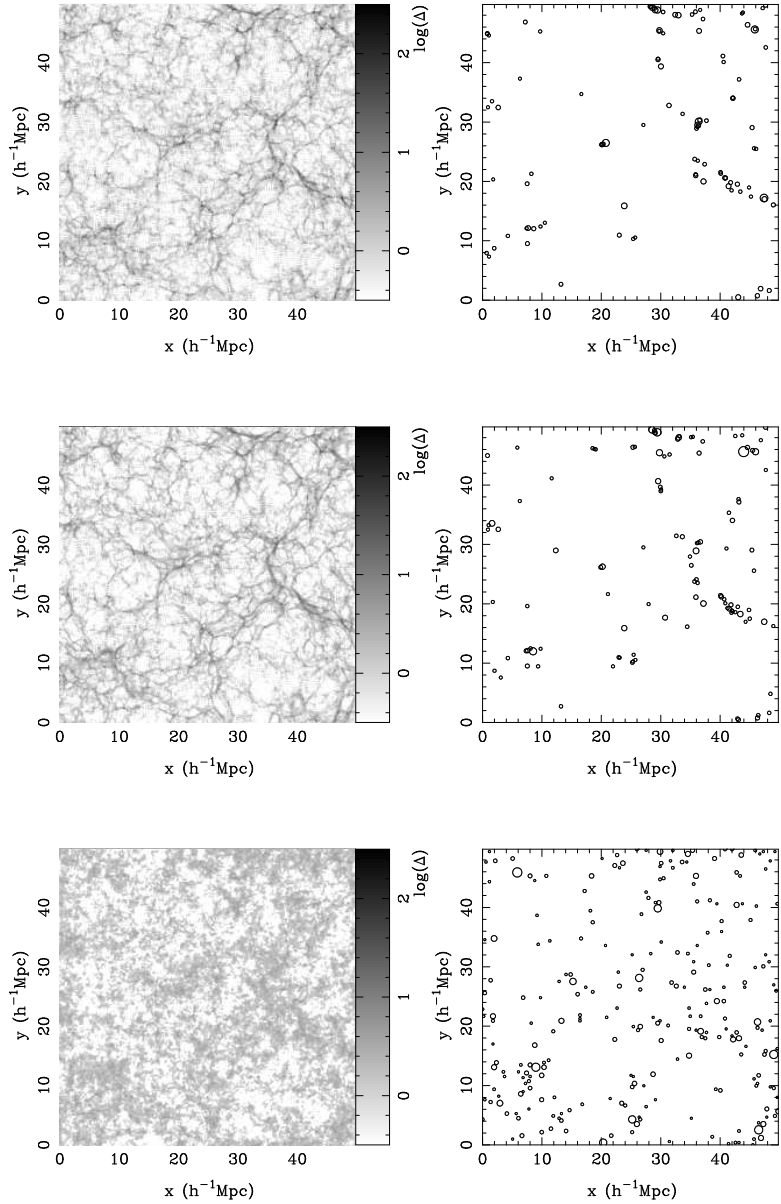}}}
\caption{The density field (left panel) and the location of collapsed
haloes  (right panel) at $z=6$ obtained from the three methods,
namely,  ``N-body + FoF'' (top panel), ``ZA + FoF'' (middle panel) and
``PS + ES'' (bottom panel).  The thickness of the slice shown is $0.2
h^{-1}$Mpc.}
\label{fig:plotdensity}
\end{figure*}

\begin{figure*}
\rotatebox{270}{\resizebox{0.3\textwidth}{!}{\includegraphics{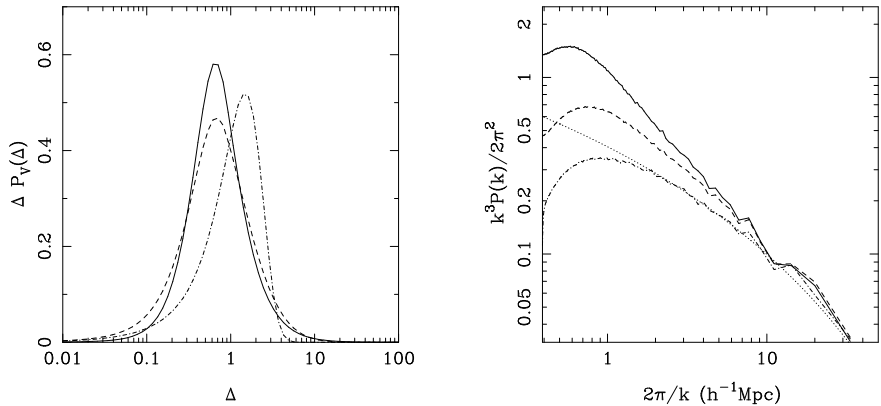}}}
\caption{Left panel: the volume-averaged probability distribution of
the density field $P_V(\Delta)$ obtained from N-body simulations
(solid curve),  Zel'dovich approximation (dashed curve) and gaussian
random field (dot-dashed curve) respectively. Right panel: the power
spectrum of density fluctuations obtained by the same three methods.}
\label{fig:densitydist}
\end{figure*}

In this Appendix, we compare  three different ways of generating the density
field and locating haloes within the simulation volume. 
Dark matter haloes were identified for density distributions 
with identical  initial conditions within a
simulation  box of comoving length 50 $h^{-1}$ Mpc with $256^3$
particles, giving a mass resolution of $5.4 \times 10^8 h^{-1}
M_{\odot}$. For definiteness, we  concentrate our comparison on 
$z=6$ (which is the fiducial redshift of study throughout the paper).

\begin{itemize}

\item{\it N-body + FoF:}  In this approach,  the dark matter density
field  is generated by  running a full N-body simulation (with 
GadgetII) and then a standard Friends-of-friends (FoF) algorithm with
linking length $b \approx 0.2$ times the mean inter-particle separation
is applied to find the haloes. Typically, one is able to  identify
haloes as small as $\sim 20$ times the mass resolution which are
consistent with  theoretical predictions of halo mass
function. This is most accurate method to  obtain
the spatial distribution of dark matter haloes.  
The disadvantage is that  in order to achieve the  dynamic range required 
for studying reionization is generally computationally  expensive (both in
terms of CPU time and memory). The density field and the location of
dark matter haloes obtained in this way are  shown in the top panels
of Figure \ref{fig:plotdensity}.

\item{\it ZA + FoF:} An alternate method of generating the density
field is the  Zel'dovich approximation. In this case, we have 
generated the density field at a given redshift by displacing
the particles from their initial positions using the linear velocity
field.  This procedure is significantly  less computationally expensive than a
N-body simulation and nevertheless gives give a reasonable representation of the
density field at high redshifts. The location and mass of the  haloes
was then obtained with  FoF halo finder  with
a variable linking length with $b \approx 0.3-0.35$. The 
detailed internal structure of the haloes is not correct in this case 
(the density profiles of the haloes is generally much more diffuse
and the halo particle may even not  not be bound). However, these  details
are not important  for our work here where we want to investigate
qualitatively the topology of reionization.  
The density field and the location of the haloes obtained in this way
are shown in the
middle panels of Figure \ref{fig:plotdensity}. One  immediately
appreciates that the visual impression of both the density structure
and halo field generated by this approach is very similar to the
previous one, the differences being rather  minor.

\item{\it GRF + ES:} The third method we have explored is evolving
the initial Gaussian random field (GRF) linearly (i.e., multiplying by
the appropriate growth factor) and applying the excursion set (ES)
formalism to identify the haloes. The advantage in this case is that
the formalism is computationally very cheap  and can identify
haloes as small as the mass resolution of the box. The disadvantage is
that the linear density field does not necessarily capture  the true
density distribution which is a serious problem for the analyses
presented here. The results obtained by this approach are
shown in the bottom panels of Figure \ref{fig:plotdensity}. It is
immediately apparent that the density structure is drastically different
from the previous two approaches with no apparent filamentary networks
visible. The same is true for the location of the  haloes (though it
should be mentioned that the number of haloes identified are much
larger than the previous methods as one can locate smaller
haloes). A better match with the simulations can be achieved if both the
densities and halo positions are adjusted using the Zel'dovich approximation
\cite{zlm+07,mf07}; however it is not clear how well the density peaks would correspond to
halo locations if both are displaced independently. Since a reasonable representation of the density field and
location of the haloes are vital for our work here, this very simple 
computationally least expensive scheme is unfortunately not  appropriate for this work.

The fact that the ``ZA + FoF'' method gives  a reasonable
approximation of the density and halo field can also be seen
quantitatively from Figure \ref{fig:densitydist} where we have
plotted the volume-weighted density distribution $P_V(\Delta)$ (left
panel) and the power spectrum of density fluctuations $P(k)$ (right
panel) for the three methods. The density
distribution obtained with  the ``ZA + FoF'' method (dashed curve)
closely resembles that obtained with  ``Nbody + FoF'' (solid curve),
which is quite different from the gaussian distribution (dot-dashed
curve) obtained with the ``GRF + ES'' method. Similarly, the plots of
the power spectrum shows that the ``GRF + ES'' method deviates from the
``Nbody + FoF'' at scales $\sim 10 h^{-1}$ Mpc, while the ``ZA + FoF''
method  is reasonable down to scales of a  few $h^{-1}$ Mpc. At smaller
scales, the ``Nbody + FoF'' method generates more power than the other
two cases due to a correct treatment of  non-linearities.  It appears
thus  fair to say that the ``ZA + FoF'' method is  a good approximation for scales
$\gtrsim 1 h^{-1}$ Mpc, which should be  sufficient for generating the
ionization maps in this work.

\end{itemize}

\section{Numerical convergence}
\label{app:numres}

\begin{figure}
\rotatebox{270}{\resizebox{0.45\textwidth}{!}{\includegraphics{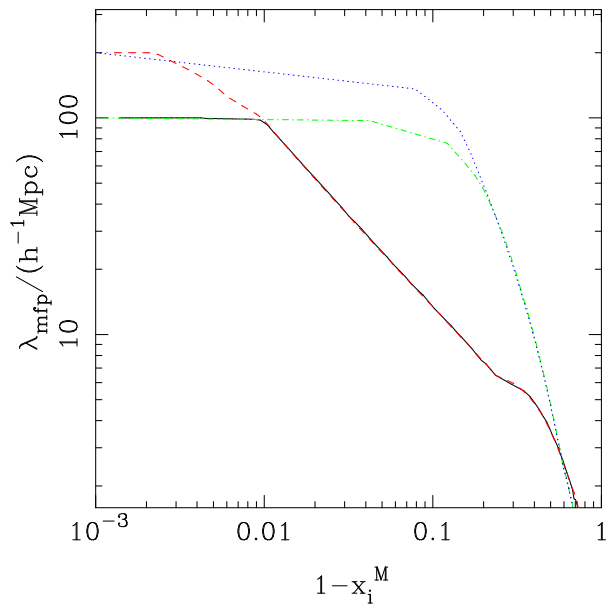}}}
\caption{The effect of box size on the photon mean free path for
  models  HR and IR-1.0. The solid (dot-dashed) and the dashed
  (dotted) curves represent the results for the  
100$h^{-1}$ Mpc and
200$h^{-1}$ Mpc box, respectively,  for the IR-1.0 (HR) model.}
\label{fig:plot_compare_mfp_boxsize}
\end{figure}

\begin{figure}
\rotatebox{270}{\resizebox{0.45\textwidth}{!}{\includegraphics{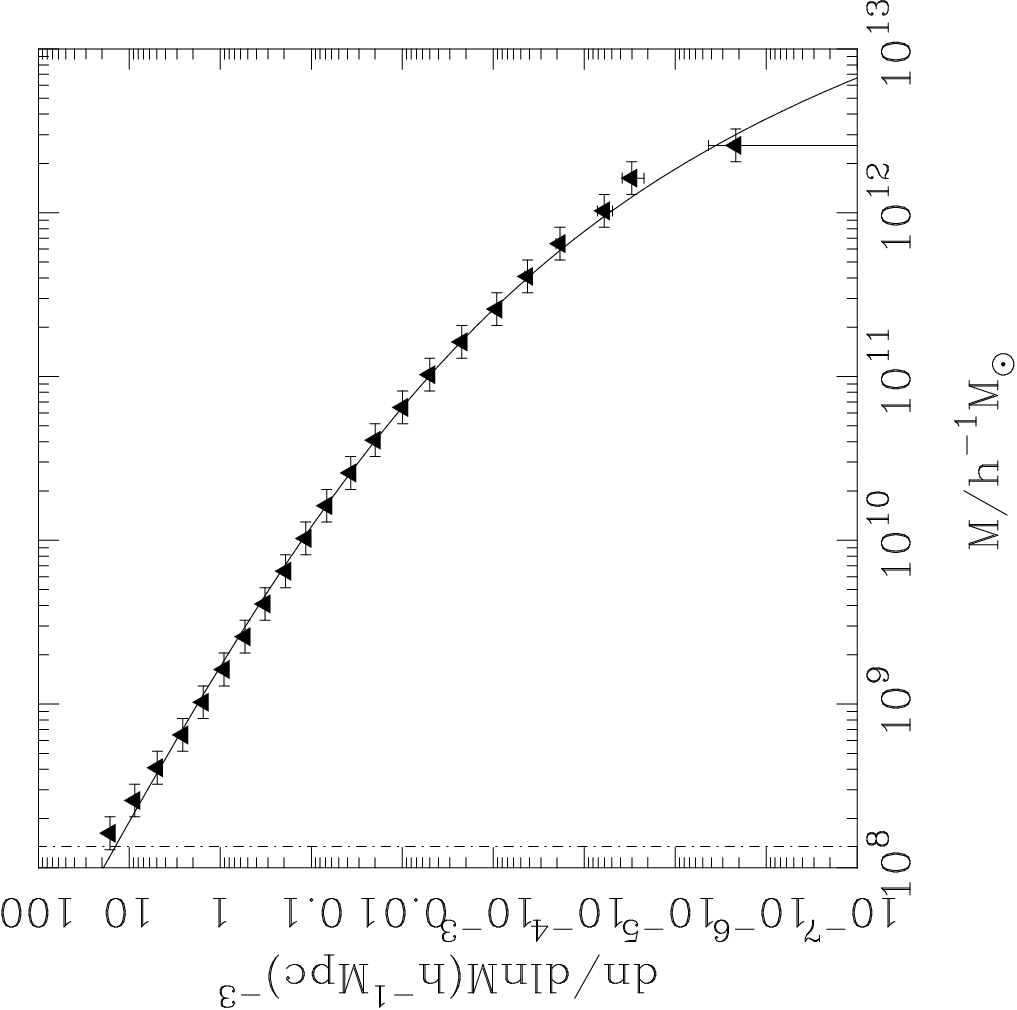}}}
\caption{The halo mass function at $z=6$ for the high-resolution
simulation. The points with errorbars show the results from our
simulation; the vertical errors correspond to the statistical
uncertainties while the horizontal errors denote the bin size. The
solid curve is the theoretical mass function of Sheth \& Tormen
(2002),  with the fitting function adopted from Jenkins et
al. (2001).}
\label{fig:haloes-multb-large}
\end{figure}

\begin{figure}
\rotatebox{270}{\resizebox{0.45\textwidth}{!}{\includegraphics{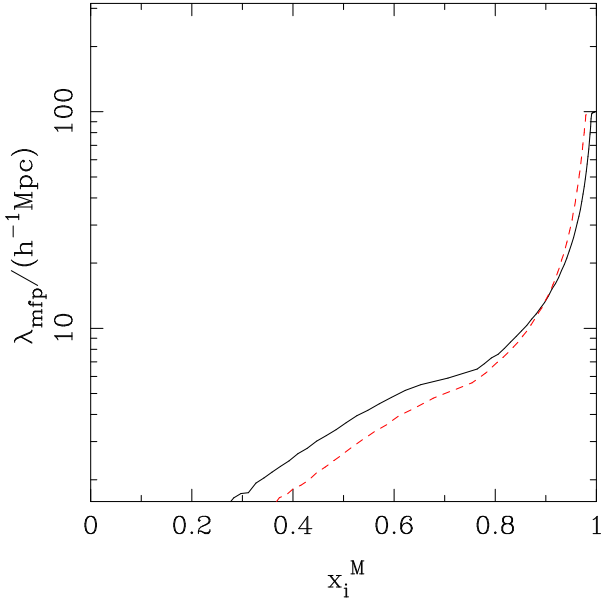}}}
\caption{Effect of resolution on the photon mean free path for model
IR-1.0. The solid curve represents the lower resolution simulation with 
$1000^3$ particles, while the dashed curve is for the higher resolution
simulation with $2000^3$ particles. The box size  in both  cases is 100$h^{-1}$ Mpc.}
\label{fig:plot_compare_mfp_resol}
\end{figure}

In this appendix, we discuss the effects of limited box size and mass
resolution on our results.  For simplicity, we shall keep our
discussion focussed on the models HR and IR-1.0.

In order to study the effect of box size, we have run a simulation
with a box of length 200 $h^{-1}$ Mpc (comoving) with $2000^3$
particles, thus giving the same mass resolution as our fiducial box.
We find that the effect on quantities like ionized fraction
$Q_i(\Delta)$ and  the distribution of 21cm brightness temperature
$P_M(\Delta_{21})$  is negligible for all models.  The only
significant effect of a larger box size concerns  the evolution of the
photon mean free path $\lambda_{\rm mfp}$ (which is shown in  Figure
\ref{fig:plot_compare_mfp_boxsize}) and, to some extent,  the  21cm
power spectrum $\Delta_{21}^2(k)$. 

For model  IR-1.0, we find no significant effect of the limited box
size  on the shape or amplitude of $\Delta_{21}^2(k)$ other that  we
are able to probe larger scales  with a larger box size.  The mean
free path  $\lambda_{\rm mfp}$ is not affected by the limited box size
for scales smaller than the box as can be seen by comparing the solid
and dashed curves in  Figure \ref{fig:plot_compare_mfp_boxsize}).
However, with our fiducial box size of 100 $h^{-1}$ Mpc, it is not
possible to probe the IGM when the mass-averaged neutral fraction $1 -
x_i^M < 0.01$. If the box size is doubled to 200 $h^{-1}$ Mpc, we are
able to probe  a much smaller neutral fraction $1 - x_i^M <
0.002$. This confirms the result that larger boxes are essential when
reionization enters its final stages.

The requirement for larger box sizes is more apparent for model  HR, 
where we find that the limited box size affects the value of
$\lambda_{\rm mfp}$ for scales about half the box size (dotted and
dot-dashed curves in Figure \ref{fig:plot_compare_mfp_boxsize}).  In
fact, we find that a box size of as large as 100 $h^{-1}$ Mpc is 
only sufficient for neutral fractions $1 - x_i^M > 0.25$.  This is
not surprising as the HR model tends to produce large ionized regions
whose growth can be affected seriously with a limited box size.  We
come to similar conclusions when studying the 21cm power spectrum. 
However the differences are not as statistically  significant as the
number of points which are neutral decreases during the  late stages of
reionization.

Finally, we present the effect of numerical resolution on our
analyses.  For this purpose, we have run a simulation box  of length
100 $h^{-1}$ Mpc (comoving) with $2000^3$ particles, which gives a
mass resolution of $M_{\rm part} = 9.02 \times 10^6 h^{-1} M_{\odot}$.
Applying the FoF method with adaptive linking length on this
distribution,  we are able to locate haloes as small as $9.02 \times
10^7 h^{-1} M_{\odot}$, thus achieving sensitivities corresponding to
haloes able to cool via atomic transitions. The mass function of
haloes at $z=6$ for this high-resolution simulation is shown in Figure
\ref{fig:haloes-multb-large};  we have also shown the corresponding
theoretical mass function \cite{jfw++01} for comparison.  The halo
mass function agrees now very well with the theoretical expectation
for an even  larger dynamic range.

In Section \ref{sec:results} we have shown that the spatial distribution of
sources of ionizing radiation have a huge effect on the
ionization fields. Thus, it is
naturally expected that the ionization maps would be very different
for a high resolution box if we include all the low-mass
sources. However, our main concern is to study the resolution effects
for an identical source distribution is identical. Keeping that in mind,
we include only sources with $M > 10^9 h^{-1} M_{\odot}$ so that the
source distribution is statistically identical to that in our fiducial
box. For the high resolution box, 
we smooth the density field to a grid-size of $0.5 h^{-1}$ Mpc
corresponding to $200^3$ grid points in the box.

The main effect of the resolution enters into our results through the
recombination rate. Since it is dependent on the local density,  we
find that the rate is higher when we include high resolution  (i.e.,
high density) pixels in the analysis. We would thus expect, for example,
that the mean free path is smaller  in the high
resolution simulation  (even when the source distribution is statistically
similar). That is indeed the case as is shown in Figure
\ref{fig:plot_compare_mfp_resol} where we have compared the high
resolution simulation  with the fiducial simulation model  IR-1.0. 
At the very late stages of reionization, however, the mean free path
in the two cases is similar. In fact, at large ionized mass fraction, 
the only structures to remain neutral have  intermediate densities,
which should be  equally well probed by
the two simulations with different resolution.

\end{document}